\documentclass[11pt,a4paper]{article}
\pdfoutput=1
\usepackage{jcappub}
\usepackage{rotating}
\usepackage{graphicx,epsfig}
\usepackage{amsmath}
\usepackage{amssymb}
\usepackage{subfigure,time}
\usepackage{relsize}
\usepackage[utf8]{inputenc}

\title{Magnetically charged black holes from non-linear electrodynamics and the Event Horizon Telescope}

\author[a]{Alireza Allahyari,}
\author[a]{Mohsen Khodadi,}
\author[b]{Sunny Vagnozzi,}
\author[c]{and David F. Mota}

\affiliation[a]{School of Astronomy, Institute for Research in Fundamental Sciences (IPM), Niavaran Square, P.O. Box 19395-5746, Tehran, Iran}
\affiliation[b]{Kavli Institute for Cosmology (KICC) and Institute of Astronomy, University of Cambridge, Madingley Road, Cambridge CB3 0HA, United Kingdom}
\affiliation[c]{Institute of Theoretical Astrophysics, University of Oslo, P.O. Box 1029 Blindern, N-0315 Oslo, Norway}

\emailAdd{alireza.al@ipm.ir}
\emailAdd{m.khodadi@ipm.ir}
\emailAdd{sunny.vagnozzi@ast.cam.ac.uk}
\emailAdd{d.f.mota@astro.uio.no}

\abstract{Non-linear electrodynamics (NLED) theories are well-motivated extensions of QED in the strong field regime, and have long been studied in the search for regular black hole (BH) solutions. We consider two well-studied and well-motivated NLED models coupled to General Relativity: the Euler-Heisenberg model and the Bronnikov model. After carefully accounting for the effective geometry induced by the NLED corrections, we determine the shadows of BHs within these two models. We then compare these to the shadow of the supermassive BH M87* recently imaged by the Event Horizon Telescope collaboration. In doing so, we are able to extract upper limits on the black hole magnetic charge, thus providing novel constraints on fundamental physics from this new extraordinary probe.}

\begin{document}
\maketitle
\flushbottom

\section{Introduction}
\label{sec:intro}

Black holes (BHs) are exceptionally remarkable regions of spacetime. They are defined by their event horizon, a one-way causal spacetime boundary beyond which even light is unable to escape. A generic prediction of General Relativity (GR)~\cite{Einstein:1916vd,Schwarzschild:1916uq,Penrose:1964wq}, BHs are understood to constitute the end state of gravitational collapse of matter. Moreover, BHs might provide the key towards the dream of unifying General Relativity (GR) and Quantum Mechanics~\cite{Hawking:1976ra,Giddings:2017jts} (see also~\cite{Giddings:2019jwy}). There is no doubt that a better understanding of BHs will lead to a deeper understanding of gravity at energy scales unaccessible to us from Earth.

From the observational point of view, BHs appear in a wide variety of astrophysical environments, and there is wealth of direct or indirect evidence pointing towards the existence of supermassive black holes (SMBHs), with masses as large as $10^{10}\,M_{\odot}$. In fact, it is believed that SMBHs reside in the centre of most sufficiently massive galaxies, including ours~\cite{LyndenBell:1969yx,Kormendy:1995er}, and that they power so-called active galactic nuclei, extremely luminous central regions of galaxies that oftentimes outshine the rest of the galaxies themselves. See e.g.~\cite{Bambi:2019xzp} for an up-to-date comprehensive review on astrophysical BHs.

Due to the combination of a photon sphere (where photons travel along unstable circular orbits) and gravitational lensing of photons, an accreting BH surrounded by a geometrically thick, optically thin emission region will lead to the appearance of a so-called shadow, in combination with a bright emission ring~\cite{Luminet:1979nyg,Lu:2014zja,Cunha:2018acu,Gralla:2019xty,Narayan:2019imo}. The BH shadow represents the interior of the so-called ``apparent boundary'' or ``critical curve'', the latter being such that a light ray belonging to the critical curve asymptotically approaches a bound photon orbit when traced from a distant observer back to the BH. In other words, the BH shadow represents the closed curve on the sky separating capture orbits and scattering orbits. See e.g.~\cite{Dokuchaev:2019jqq} for a recent review on BH shadows. Very long baseline interferometry (VLBI) surveys, wherein signals from various astronomical radio sources are collected at multiple radio telescopes on Earth, effectively emulating a single huge telescope with size given by the maximum separation between the individual telescopes, are expected to be able to detect the shadows of SMBHs~\cite{Falcke:1999pj}.

VLBI interferometry in the context of BH shadows has become a reality through the Event Horizon Telescope (EHT) collaboration, an Earth-wide radio telescope array observing at $1.3\,{\rm mm}$ wavelength with a theoretical diffraction-limited resolution of $25\,\mu{\rm as}$~\cite{Doeleman:2009te}. In April 2019, the EHT collaboration announced the detection of the shadow of M87*, the SMBH residing at the center of the elliptical galaxy Messier 87, in a series of by now seminal papers~\cite{Akiyama:2019cqa,Akiyama:2019brx,Akiyama:2019sww,Akiyama:2019bqs,Akiyama:2019fyp,Akiyama:2019eap}. Broadly speaking, the image of M87*'s shadow appears to be consistent with that of a Kerr BH in GR, but further improvements are required before more can be said. At any rate, BH shadows appear to be an extremely promising arena for testing deviations from GR~\cite{Johannsen:2010ru,Loeb:2013lfa,Johannsen:2015hib,Johannsen:2015mdd,Psaltis:2018xkc}, and in particular violations of the no-hair theorem~\cite{Israel:1967wq,Israel:1967za,Carter:1971zc}. In fact, following the announcement of the EHT detection, several works examined the possibility of extracting valuable information from M87*'s shadow, ranging from properties of the BH itself such as its mass and spin to novel constraints on fundamental physics, see e.g.~\cite{Moffat:2019uxp,Nokhrina:2019sxv,Abdikamalov:2019ztb,Held:2019xde,Wei:2019pjf,Shaikh:2019fpu,Tamburini:2019vrf,Davoudiasl:2019nlo,Ovgun:2019yor,Bambi:2019tjh,Nemmen:2019idv,Churilova:2019jqx,Safarzadeh:2019imq,Firouzjaee:2019aij,Konoplya:2019nzp,Kawashima:2019ljv,Contreras:2019nih,Bar:2019pnz,Jusufi:2019nrn,Vagnozzi:2019apd,Banerjee:2019cjk,Roy:2019esk,Ali:2019khp,Long:2019nox,Zhu:2019ura,Contreras:2019cmf,Dokuchaev:2019pcx,Qi:2019zdk,Wang:2019tto,Konoplya:2019goy,Roy:2019hqf,Pavlovic:2019rim,Biswas:2019gia,Wang:2019skw,Nalewajko:2019mxh,Tian:2019yhn,Cunha:2019ikd,Banerjee:2019nnj,Shaikh:2019hbm,Vrba:2019vqh,Kumar:2019ghw}.

In GR, continuous gravitational collapse appears to lead to the inevitable but somewhat undesirable existence of singularities~\cite{Penrose:1964wq,Hawking:1969sw,Senovilla:2018aav}. While the cosmic censorship conjecture states that all singularities of gravitational collapse should be hidden behind event horizons of BHs, and hence should not be ``naked''~\cite{Penrose:1969pc,Wald:1997wa}, it is nonetheless desirable to find solutions which avoid singularities altogether. Thus, a great deal of attention has been paid to the search for regular BH solutions, starting from the early work of Bardeen~\cite{Bardeen:1968ghw}, and working by either modifying the gravity sector or seeking (typically exotic) matter content which can regularize the central singularity. For an incomplete list of works in this direction, see e.g.~\cite{Borde:1996df,AyonBeato:1998ub,AyonBeato:1999rg,Bronnikov:2000yz,bur1,Hayward:2005gi,Bronnikov:2005gm,Iso:2006ut,Berej:2006cc,bur3,bur2,Bronnikov:2012ch,Rinaldi:2012vy,Bambi:2013ufa,Sebastiani:2013fsa,Toshmatov:2014nya,Johannsen:2015pca,Myrzakulov:2015sea,Myrzakulov:2015qaa,Myrzakulov:2015kda,Abdujabbarov:2016hnw,Fan:2016hvf,Sebastiani:2016ras,Toshmatov:2017zpr,Chinaglia:2017uqd,Chinaglia:2017wih,Colleaux:2017ibe,Jusufi:2018jof,Chinaglia:2018gvf,Ovgun:2019wej,Han:2019lfs,Rodrigues:2019xrc,Panotopoulos:2019qjk,Jusufi:2019caq,Gorji:2019hog}. See~\cite{Stuchlik:2014qja,Schee:2015nua,Schee:2016mjd,Stuchlik:2019uvf,Schee:2019gki} for important works regarding observational signatures of such BHs.

A very attractive class of models emerging in the search for regular BH solutions is non-linear electrodynamics (NLED). A classical example in this sense is Born-Infeld (BI) electrodynamics~\cite{Born:1934gh}, first introduced as a classical solution to the electron self-energy problem. Regardless of their utility in the search for regular BH solutions, NLED models are in any case relevant when taking into account loop corrections to quantum electrodynamics (QED), necessary when one wants to describe the strong-field regime of the electromagnetic field, for instance when tackling the issue of the self-energy problem of a point charged particle. One particularly relevant example in this sense is Euler-Heisenberg (EH) NLED, whose action is given by the effective action of QED after taking into account one-loop corrections~\cite{Heisenberg:1935qt}, and which sees the appearance of two relativistic invariants constructed out of the electromagnetic field-strength tensor. Another important example is that of the Bronnikov NLED model~\cite{Bronnikov:2000vy}, which instead only sees the appearance of one of the previous two relativistic invariants, and wherein regular BH solutions exist provided they only carry magnetic and not electric charge.

We wish to remark at this point that, contrary to popular belief, NLED models such as the EH model are not exotic. In fact, they are the inevitable extension of the better-known theory of electromagnetic interactions, QED (and, by extension, of Maxwell's equations), in the high-intensity regime, which is relevant for several studies~\cite{Stehle:1966wii}. Moreover, it is known that NLED models frequently appear in the low-energy limit of several string theories or supersymmetric theories. For instance, it is known that BI NLED appears as the low-energy effective description of world-volume gauge fields on
D-branes~\cite{Fradkin:1985qd,Tseytlin:1986ti}. On the other hand, the EH NLED is effectively the low-energy limit of BI NLED, and is known to well approximate the supersymmetric action of minimally coupled spin-$1/2$ and spin-$0$ particles~\cite{Bern:1993tz,Dunne:2004nc,Jacobson:2018kso}.

Black hole solutions in NLED models were studied in a wide number of papers. For instance, within BI NLED, an incomplete list of works studying BH solutions can be found in~\cite{Dey:2004yt,Cai:2004eh,Stefanov:2007qw,HabibMazharimousavi:2008dm,Stefanov:2009qd,Ghodsi:2010ev,Allahverdizadeh:2013oha,Allaverdizadeh:2013rua,Olmo:2013gqa,Wu:2016hry,Bambi:2016xme,Chen:2017ify,Boehmer:2019uxv,Kumar:2019zbp,Li:2019qbw}. Similar studies within the context of Euler-Heisenberg non-linear electrodynamics can instead be found in~\cite{Bronnikov:1979ex,Yajima:2000kw,Bronnikov:2000vy,Stefanov:2007zza,Corda:2009xd,Ruffini:2013hia,Hendi:2014xia,Breton:2016mqh,Maceda:2018zim,Olvera:2019unw}. Of particular interest to us is~\cite{Bronnikov:1979ex}, where it was argued that regular BH solutions are not possible for configurations with non-zero electric charge, under the assumptions of static spherical symmetry. Later on~\cite{Bronnikov:2000vy} showed that this result persists even for dyonic configurations, where both non-zero electric and magnetic charges are present. However, still~\cite{Bronnikov:2000vy} showed that this ceases to be the case if one considers a configuration with a pure magnetic charge. Notice that, while the field-strength $F^{\mu \nu}$ can be infinite at the centre, in the same point, wherein the force applied to test particles vanishes, both the energy-momentum tensor and the metric are at least $C^2$. Such BH solutions were studied in detail in~\cite{Bronnikov:2000vy}, and shown to have interesting properties.~\footnote{See also~\cite{Kruglov:2017fck,Kruglov:2017mpj,Kruglov:2017xmb,Kruglov:2018rrm,Kruglov:2018lct} for further work on magnetically charged BHs in non-linear electrodynamics.}

Our goal in this work is three-fold. First of all, we want to investigate novel BH solutions in the EH and Bronnikov NLED models coupled to GR, which we refer to as Einstein-Euler-Heisenberg and Einstein-Bronnikov models respectively. We focus on BH solutions carrying magnetic charge, considering both regular and non-regular BHs. The next point we want to address is to carefully compute the shadows of the resulting BHs. This is important because it has been shown in~\cite{Novello:1999pg} that due to non-linear electrodynamics, photons actually propagate along geodesics that are no longer the geodesics of the original spacetime, but in the so-called \textit{effective geometry}. This fact has not been appreciated sufficiently in the literature, but is crucial when computing the resulting BH shadow, given the importance of null geodesics in the analysis of the latter. Finally, we wish to compare the shadows we find with the shadow of M87* detected by the EHT collaboration, and possibly set novel and valuable limits on the fundamental parameters of the two NLED models we are considering.

The rest of this work is then organized as follows. We begin in Sec.~\ref{sec:shadownonregular} by discussing Einstein-Euler-Heisenberg gravity. We then consider non-regular magnetically charged BHs in the theory, and compute the resulting shadows. We then do the same in Sec.~\ref{sec:shadowregular} for regular magnetically charged BHs in Einstein-Bronnikov gravity. In Sec.~\ref{sec:ehtshadow} we then compare the resulting shadows to the shadow of M87* detected by the Event Horizon Telescope collaboration, and study whether we can use the latter to set constraints on the free parameters of the underlying non-linear electrodynamics models, and in particular on the magnetic charge $Q_m$. Finally, we provide concluding remarks in Sec.~\ref{sec:conclusions}. Throughout this work, we use Planck units with $c=\hbar=G=1$.

\section{Shadows of non-regular magnetic Einstein-Euler-Heisenberg black holes}
\label{sec:shadownonregular}

In this section, we will begin by studying the motions of photons in a non-regular spacetime arising as a solution to the Einstein-Euler-Heisenberg theory of gravity. We will then use these results to determine the shadows of Einstein-Euler-Heisenberg non-regular BHs possessing a magnetic charge. The crucial point is that due to the non-linear electrodynamics, photons follow null geodesics of an induced effective geometry rather than those of the original background spacetime.~\footnote{For a recent study on the thermodynamics of non-linear magnetically charged BHs, see e.g.~\cite{Ndongmo:2019ywh}.}

\subsection{Static magnetically charged black holes in Einstein-Euler-Heisenberg gravity}
\label{subsec:static}

As we explained earlier, Euler-Heisenberg non-linear electrodynamics is the low-energy limit of Born-Infeld electrodynamics. In EH NLED, the standard Maxwell Lagrangian is modified to:
\begin{eqnarray}
\mathcal{L}(U,W)=-\frac{1}{4}U+\frac{\mu}{4}(U^2+\frac{7}{4}W^2)\,,
\label{LFP}
\end{eqnarray}
where the two relativistic invariants $U$ and $W$ are constructed from the electromagnetic field-strength tensor $F_{\alpha\beta}$ and its dual $F^{\star}_{\alpha\beta}$ as follows:
\begin{eqnarray}
U=F^{\alpha\beta}F_{\alpha\beta}\,, \quad W=F^{\alpha\beta}F_{\alpha\beta}^{\star}\,, \quad F^{\star}_{ \alpha\beta}=\frac{1}{2}\epsilon_{\alpha\beta\mu\nu}F^{\mu\nu}\,, \quad F_{\mu \nu} = \partial_{\mu}A_{\nu}-\partial_{\nu}A_{\mu}\,,
\label{fg}
\end{eqnarray}
with $\star$ and $\epsilon_{\alpha\beta\mu\nu}$ representing respectively the Hodge dual operator and the completely antisymmetric Levi-Civita tensor obeying $\epsilon_{\alpha\beta\mu\nu}\epsilon^{\alpha\beta\mu\nu}=-4!$, whereas $A_{\mu}$ is the electromagnetic gauge field. Note that the modified action given in Eq.~(\ref{LFP}) deviates from the standard Maxwell Lagrangian via the positive dimensionless constant $\mu$.

When coupling GR to EH NLED, we refer to the resulting theory as Einstein-Euler-Heisenberg (EEH) gravity, with action given by:
\begin{eqnarray}
S=\frac{1}{16\pi}\int d^4x\,\sqrt{-g}  \left [ R+ 4\mathcal{L}(U,W) \right ]\,.
\label{L1}
\end{eqnarray}
The gravitational field equations and the energy-momentum tensor sourced by EH NLED are found upon taking the variation of Eq.~(\ref{L1}) with respect to the metric tensor:
\begin{eqnarray}
G_{\mu\nu}=8\pi T_{\mu\nu} =\frac{1}{2}g_{\mu\nu}\mathcal{L}+(2-4\mu U)\, {F_{\mu}}^{\alpha} \, F_{\alpha\nu}-56\mu W\epsilon_{\mu\sigma\rho\gamma} {F}^{\sigma\rho} \, F_{\nu}^{\gamma}\,.
\label{T}
\end{eqnarray}
Henceforth, in order to solve the field equations, we shall make the ansatz of a static spherically symmetric (SSS) metric, whose (squared) line element $ds^2_{\rm EEH}$ is given by:
\begin{eqnarray}
ds^2_{\rm EEH}=-f_{\rm EEH}(r)dt^2+\frac{dr^2}{f_{\rm EEH}(r)}+r^2d\theta^2+r^2\sin^2\theta d\varphi^2\,,
\label{ds}
\end{eqnarray}
and is characterized by the metric function $f_{\rm EEH}(r)$, which is solely a function of the radial coordinate.

As discussed earlier, we shall consider BHs carrying magnetic but not electric charge. This is achieved by choosing a purely magnetic configuration for the electromagnetic gauge field $A_{\mu}$, given by:
\begin{eqnarray}
A_\mu=Q_m\cos\theta \delta_{\mu}^{\varphi}\,,
\label{gauge}
\end{eqnarray}
where $Q_m$ is the magnetic charge.
After solving the gravitational field equations [Eq.~(\ref{T})] for the effective Einstein-Euler-Heisenberg action given by Eq.~(\ref{L1}), from the $G^t_t$ component we find:
\begin{align}
\frac{f'_{\rm EHH}}{r}+\frac{f_{\rm EHH}}{r^2}-\frac{2\mu Q_m}{r^8}+\frac{Q_m^2}{r^4}-\frac{1}{r^2}=0\,,
\end{align}
whereas from the $G^\theta_\theta$ component we find:
\begin{align}
f''_{\rm EHH}+\frac{2f'_{\rm EHH}}{r}+\frac{1}{r^4}\left(12\mu Q_m^4-2 Q_m^2 \right) =0\,.
\end{align}
The metric function $f_{\rm EEH}(r)$ appearing in Eq.~(\ref{ds}) is found to be:
\begin{eqnarray}
f_{\rm EEH}(r)=\bigg(1-\frac{2M}{r}+\frac{Q_m^2}{r^2}-\frac{2\mu}{5}\frac{Q_m^4}{r^6}\bigg)\,.
\label{Nr}
\end{eqnarray}
As we could have expected, the static-charged black hole solution with squared line element given by Eq.~(\ref{ds}) is parametrized by the black hole mass $M$, as well as the magnetic charge $Q_m$. It is worth noting that the authors of~\cite{Maceda:2018zim} have found the electric counterpart of Eq.~(\ref{Nr}), which unsurprisingly has an analogous form to the solution we found. One can see that by setting $\mu\rightarrow0$, the standard Reissner-Nordstr\"{o}m (RN) squared line-element for a rotating charged BH is recovered. Finding exact expressions for the relevant horizons from the metric function Eq.~(\ref{Nr}) is not straightforward. However, using \textit{Descartes' rule of signs}, which provides us the number of real zeros of an arbitrary polynomial function, we can guess that the polynomial in question will have one or three positive roots, which will be related to the relevant horizon(s). Henceforth, we shall work in units of mass setting $M=1$, or equivalently rescale all dimensionful quantities by the appropriate power of $M$.

\subsection{Effective geometry induced by non-linear Euler-Heisenberg electrodynamics}
\label{subsec:effectivegeometry1}

In this section, we will use Novello's method \cite{Novello:1999pg,DeLorenci:2000yh} to derive the effective geometry induced by Euler-Heisenberg non-linear electrodynamics effects, which alter the background geometry along the null geodesics of which photons would usually propagate. Inspired by the Lagrangian given in Eq.~(\ref{LFP}), we consider now a general non-linear electrodynamics Lagrangian given by $\mathcal{L}=\mathcal{L}(U,\,W)$, which again depends on both the relativistic invariants $U$ and $W$. Using the least action principle, we find the equations of motion to be:
\begin{eqnarray}
\partial_{\alpha}\,\left(\mathcal{L}_{U}F^{\alpha\beta}+\mathcal{L}_{W}F ^{ \star \alpha\beta}\right) = 0\,, \quad \mathcal{L}_{U,W}=\frac{d\mathcal{L}_{U,W}}{dU(W)}\,.
\label{field}
\end{eqnarray}
Let us consider the constant phase surface $\Sigma$. We require that electromagnetic fields are continuous across this surface and their derivative is discontinuous. By imposing the conditions $[F_{\alpha\beta}]_\Sigma=0$ and $[\partial_\sigma F_{\alpha\beta}]_\Sigma=f_{\alpha\beta}k_\sigma$ on the surface of discontinuity $\Sigma$, where $[F_{\alpha\beta}]_\Sigma=F_{\alpha\beta}^{+}-F_{\alpha\beta}^-$,  the first of the two equations in Eq.~(\ref{field}) becomes:
\begin{eqnarray}
\bigg(\mathcal{L}_U\,f^{\alpha\beta}+a_1\,F^{\alpha\beta}+a_2\,F ^{\star \alpha\beta}\bigg)\,k_\alpha = 0\,,
\label{ef}
\end{eqnarray}
where we have defined:
\begin{eqnarray}
a_1&\equiv 4\,\left ( F^{\alpha\beta} f_{\alpha\beta}\,\mathcal{L}_{UU}+ F^{\star \alpha\beta} f_{\alpha\beta}\,\mathcal{L}_{UW} \right )\,,\\
a_2&\equiv 4\,\left ( F^{\alpha\beta} f_{\alpha\beta}\,\mathcal{L}_{UW}+F^{\star \alpha\beta} f_{\alpha\beta}\,\mathcal{L}_{WW} \right )\,.
\label{eff}
\end{eqnarray}
By contracting Eq.~(\ref{ef}) with $F^\alpha{}_\mu k_\alpha$ and $F ^{\star \alpha}{}_\mu k_\alpha$, we get the following expressions respectively:
\begin{eqnarray}
\left ( F^{\alpha\beta} f_{\alpha\beta}\mathcal{L}_{U} + \frac{a_2}{4}\,W \right )\,\eta^{\mu\nu}k_{\mu}\,k_{\nu} - a_1  {F^{\nu}}_{\alpha}\, F^{\alpha\mu} k_\nu k_\mu= 0\,,
\label{gw1}
\end{eqnarray}
and:
\begin{eqnarray}
\left ( F^{\star \alpha\beta} f_{\alpha\beta}\,\mathcal{L}_{U} - \frac{a_2}{2}\,U+\frac{a_1}{4}\,W \right ) \eta^{\mu\nu}k_{\mu}\,k_{\nu} - a_2  {F^{\nu}}_{\alpha}\, F^{\alpha\mu} k_\nu k_\mu= 0\,.
\label{gw2}
\end{eqnarray}

To make progress we define $\Omega \equiv\frac{F^{\star \alpha\beta}f_{\alpha\beta}} {F^{\alpha\beta}f_{ \alpha\beta}}$. Doing so, we can then manipulate Eqs.~(\ref{gw2},\ref{gw1}) and get the following quadratic equation for $\Omega$:
\begin{eqnarray}
\Omega^{2}+\frac{\Omega_2}{\Omega_1}\Omega + \frac{\Omega_3}{\Omega_1} = 0\,,
\label{r22}
\end{eqnarray}
where we have defined:
\begin{eqnarray}
\Omega_1 & \equiv & - \mathcal{L}_U\mathcal{L}_{UW}+2F\mathcal{L}_{UW} \mathcal{L}_{WW}+W(\mathcal{L}_{WW}^2 - \mathcal{L}_{UW}^2)\,,\nonumber\\
\Omega_2 & \equiv & (\mathcal{L}_U+2W\mathcal{L}_{UW})(\mathcal{L}_{WW} -\mathcal{L}_{UU})+2U (\mathcal{L}_{UU}\mathcal{L}_{WW} + \mathcal{L}_{UW}^2)\,,\nonumber\\
\Omega_3 & \equiv & \mathcal{L}_U\mathcal{L}_{UW}+2U\mathcal{L}_{UU}\mathcal{L}_{UW}+W(\mathcal{L}_{UW}^2 - \mathcal{L}_{UU}^2)\,.
\label{Omega}
\end{eqnarray}
The quadratic equation Eq.~(\ref{r22}) has two solutions which we denote $\Omega_{\pm}$:
\begin{eqnarray}
\Omega_{\pm}= -\frac{\Omega_2}{2\Omega_1}\pm\sqrt{\left ( \frac{\Omega_2}{2\Omega_1} \right )^2-\frac{\Omega_3}{\Omega_1}}\,.
\label{Omegaa}
\end{eqnarray}
If we now factor $k_\mu k_\nu$ and insert the solutions found in Eq.~(\ref{Omegaa}) into Eqs.~(\ref{gw1},\ref{gw2}), we get to the following expression describing the motion of photons:
\begin{eqnarray}
g_{\rm eff(\pm)}^{\mu\nu}\,k_{\mu}\,k_{\nu}=0\,.
\label{g2}
\end{eqnarray}
It is clear that Eq.~(\ref{g2}) describes null geodesics, thus photon paths, on an effective spacetime with metric $g_{\rm eff(\pm)}^{\mu\nu}$ given by the following expression:
\begin{eqnarray}
g_{\rm eff(\pm)}^{\mu\nu} &=&\mathcal{L}_U\eta^{\mu\nu} - 4 \left ( \left( \mathcal{L}_{UU} + \Omega_{\scriptscriptstyle \pm} \mathcal{L}_{UW}\right)F^{\mu}\mbox{}_{\lambda} F^{\lambda\nu} + \left(\mathcal{L}_{UW} + \Omega_{\pm} \mathcal{L}_{WW} \right ) F^{\mu}\mbox{}_{\lambda}F^{*\lambda\nu} \right ) \,.
\label{geral}
\end{eqnarray}
For certain non-linear electrodynamics models, such as those considered by Bardeen~\cite{Bardeen:1968ghw} and Bronnikov~\cite{Bronnikov:2000vy}, only one relativistic invariant contributes. In this case, by neglecting $W$, Eq.~(\ref{geral}) reduces to the following simpler expression:
\begin{eqnarray}
g_{\rm eff}^{\mu\nu}=\mathcal{L}_U\eta^{\mu\nu}- 4\mathcal{L}_{UU}F ^{\mu}_{\,\,\alpha}\,.
F^{\alpha\nu}\,,
\label{geral1}
\end{eqnarray}
The above effective metric will be useful to study the non-linear electrodynamics models considered in~\cite{Bardeen:1968ghw,Bronnikov:2000vy}. It turns out to also be convenient to write the effective metric in Eq.~(\ref{geral}) in the following form, explicitly highlighting the contribution from the energy-momentum tensor appearing in Eq.~(\ref{T}) and hence the role of non-linear electrodynamics corrections:
\begin{eqnarray}
g_{\rm eff(\pm)}^{\mu\nu} = {\cal M}_{\pm}\,\eta^{\mu\nu}+{\cal N}_{\pm} \,T^{\mu\nu}\,,
\label{AB}
\end{eqnarray}
where:
\begin{eqnarray}
{\cal M}_{\scriptscriptstyle \pm} & \equiv & \mathcal{L}_U + W \left( \mathcal{L}_{UW} + \Omega_{\scriptscriptstyle \pm} \mathcal{L}_{WW} \right) + \frac{1} { \mathcal{L}_{U}} \left( \mathcal{L}_{UU} + \Omega_{\scriptscriptstyle \pm} \mathcal{L}_{UW} \right) \left( \mathcal{L} - W\mathcal{L}_{W} \right)\,,\\
{\cal N}_{\scriptscriptstyle \pm} & = & \frac{ 1 }{ \mathcal{L}_{U} } \left( \mathcal{L}_{UU}+\Omega_{\scriptscriptstyle \pm} \mathcal{L}_{UW} \right)\,.
\label{A}
\end{eqnarray}

\subsection{Shadows of Einstein-Euler-Heisenberg black holes}
\label{subsec:shadow}

Now that we have computed the effective metric for photons in EEH gravity [Eq.~(\ref{geral})], we see that the effective geometry seen by photons on the background of a magnetically charged EEH BH, following Eq.~(\ref{ds}), is given by:
\begin{eqnarray}
ds^2_{\rm EEH}=g_{\rm EEH}(r)\left( -f_{\rm EEH}(r)dt^2+\frac{dr^2}{f_{\rm EEH}(r)}\right)+h_{\rm EEH}(r)\left ( r^2 d\theta^2+r^2\sin^2\theta d\phi^2 \right )\,,
\label{effgem}
\end{eqnarray}
where we have defined:
\begin{eqnarray}
&g_{\rm EEH}\left(r \right) \equiv 1-\frac{4\mu Q_m^2}{r^4}\\
&h_{\rm EEH}\left(r \right) \equiv 1-\frac{12\mu Q_m^2}{r^4}.
\label{g}
\end{eqnarray}
An important aspect to note at this point is that the effective EEH geometry is static and spherically symmetric just as the original spacetime metric we started from. Another crucial point is that the metric functions $h_{\rm EHH}(r)$ and $g_{\rm EHH}(r)$ must be positive. Only if this holds will the underlying effective geometry not flip its signature during the photon's motion. Therefore, the range accessible to the motion of photons outside of the BH is restricted to $r>r_e$ and $r>r_{eff}=(12Q_m^2\mu)^{1/4}$, where $r_e$ is the radial coordinate of the event horizon. In the case where $r_{eff}<r_e$, the exterior region of BH is still given by $r>r_e$, while in the case where $r_{eff}>r_e$ the exterior region of the BH is given by $r>r_{eff}$. Both the previously mentioned options are in principle possible depending on the parameters $Q_m$ and $\mu$.

For photon geodesics parametrized by $x^{\mu}(\tau)$ in terms of an affine parameter $\tau$, the Lagrangian of the spacetime metric given by Eq.~(\ref{effgem}) is given by:
\begin{eqnarray}
\mathcal{L}=-f_{\rm EEH}(r)g_{\rm EEH}(r)\dot t^2+\frac{g_{\rm EEH}(r)}{f_{\rm EEH}(r)}\dot r^2 + r^2 h_{\rm EEH}(r)\dot \theta^2+r^2 h_{\rm EEH}(r) \dot\phi^2\,,
\label{eff1}
\end{eqnarray}
where the dot indicates differentiation with respect to $\tau$. Because of the assumed spherical symmetry, we can safely restrict our attention to the motion of particles along the equatorial plane, for which $\theta=\pi/2$. Therefore, the equations of motion for a null geodesic are given by:
\begin{eqnarray}
&E=f_{\rm EEH}(r) g_{\rm EEH}(r)\dot{t}\,,\label{E} \\
&L=r^2h_{\rm EEH}(r)\dot{\phi}\,,\label{L} \\
&f_{\rm EEH}(r) g_{\rm EEH}(r)\dot{t}^2-\frac{g_{\rm EEH}(r)}{f_{\rm EEH}(r)} \left(\frac{dr}{d\phi} \right)^2\dot\phi^2-r^2 h_{\rm EEH}(r)\dot\phi^2=0\,,
\label{R}
\end{eqnarray}
where by $E$ and $L$ we have denoted two of the photon's constants of motion, namely its total energy and angular momentum. If we substitute Eqs.~(\ref{E},\ref{L}) into Eq.~(\ref{R}), we can rewrite the equations of motion for a null geodesic in terms of an effective potential $V(r)$ as the following:
\begin{eqnarray}
\left ( \frac{dr}{d\phi} \right)^2=V(r)=r^4 \left(-\frac{f_{\rm EEH}(r)h_{\rm EEH}(r)}{r^2g_{\rm EEH}(r)}+\frac{E^2h_{\rm EEH}(r)^2}{L^2 g_{\rm EEH}(r)^2} \right ) \,.
\label{V}
\end{eqnarray}

To compute the BH shadow, we focus on unstable circular orbits, for which $\frac{dV(r)}{dr}=0=V(r)$. Using the effective potential given in Eq.~(\ref{V}), the condition for unstable circular orbits can be rewritten as:
\begin{eqnarray}
&&b^{-2}=\frac{E^2}{L^2}=\frac{f_{\rm EEH}(r)g_{\rm EEH}(r)}{r^2 h_{\rm EEH}(r)}\,,\label{b}\\
&&rf_{\rm EEH}(r)g_{\rm EEH}(r)h_{\rm EEH}'(r)+2 f_{\rm EEH}(r) g_{\rm EEH}(r) h_{\rm EEH}(r)-r f_{\rm EEH}(r) h_{\rm EEH}(r)g_{\rm EEH}'(r) \nonumber \\
&&-r g_{\rm EEH}(r) h_{\rm EEH}(r) f_{\rm EEH}'(r)=0\,.\label{un}
\end{eqnarray}
In the above we have defined the impact parameter $b$, whose value is given by the ratio of the photon's angular momentum and energy. The impact parameter will be directly related to the size of the shadow. Moreover, the prime denotes differentiation with respect to $r$. By plugging the relevant expressions for $f_{\rm EEH}(r)$, $g_{\rm EEH}(r)$, and $h_{\rm EEH}(r)$ found earlier into Eqs.~(\ref{b},\ref{un}), we get to the following expression for determining unstable circular orbits:
\begin{eqnarray}
&&b^{-2}=\frac{Q_m^2+ r^2- 2r}{r^4}+ \frac{\mu \left ( 38 Q_m^4 + 40 Q_m^2(r^2-2r) \right ) }{5 r^8}-\frac{\mu^2 \left ( 256 Q_m^6 + 240Q_m^4(r^2-2r) \right ) }{5 r^{12}} \nonumber \\
&&+ \frac{96\mu^3 Q_m^8}{5 r^{16}}\,,\label{b1} \\
&&5r^{14}-15r^{13}+10 Q_m^2r^{12}+\mu \left ( 80 Q_m^2r^{9}-88 Q_m^4r^8 \right ) + \mu^2 \left ( 576 Q_m^6 r^8 + 240Q_m^4(r^{10}-3r^9) \right ) \nonumber \\
&&-384 \mu^3 Q_m^8 =0\,,\label{un1}
\end{eqnarray}
It is instructive to take the limit $Q_m\rightarrow0$ in Eqs.~(\ref{un1},\ref{b1}). In this case we see that the equations describe an unstable critical curve located at $r_{c-sch} = 3$, with the relevant critical impact parameter being given by $b_{c-sch} = 3\sqrt{3}$. Both results match what is expected for a standard uncharged Schwarzschild BH.

In the case when a magnetic charge is present, Eq.~(\ref{un1}) is not exactly solvable and has to be solved numerically.
There are three metric functions that define the geometry. To have a well defined geometry we need to have $f>0$ and $h>0$. So $f=h=0$ defines the boundary region allowed by the spacetime geometry. In Tables~\ref{Nu} and \ref{Nu2} we list the numerical solutions of Eqs.~(\ref{un1}), $f=0$ and $h=0$ for various cases. We consider three different values of the magnetic charge $Q_m=0.5$, $Q_m=0.9$ and $Q_m=2$, while varying $0.1\leq\mu\leq1$. In the tables we provide the radial coordinates of the unstable critical curve $r_{ph}$ (the photon sphere), as well as the relevant horizon radius $r_e$ or $r_{eff}$, respectively. It is not difficult to numerically show that if $\mu\geq0.1$, the metric function Eq.~(\ref{Nr}) has only one positive root, meaning that here we find a single-horizon charged BH for different values of $Q_m$. At first, this may appear to be an extremal solution. However, this is not the case because $f$ does not satisfy the condition $f'(r=r_e)=0$. An interesting point that should be noted here is that there is no fundamental theoretical constraint on the value of magnetic charge in units of mass so that one may in principle have $Q_m>1$, unlike the standard RN case where the electric charge in units of mass is bounded within the interval given by $0 < Q_e \leq 1$. Generally speaking, unstable critical curves with radial coordinate less than the relevant horizon radius, \textit{i.e.} $r_{ph}<r_e$, do not contribute to the shadow because they are unable to cross the event horizon and reach an observer situated at infinity. Therefore, only unstable circular orbits with radial coordinate larger than the relevant horizon radius contribute to the shadow.

As already mentioned, for the metric function in Eq.~(\ref{Nr}), there is the possibility of three positive roots. We have found that for certain values $0<\mu<0.1$ and $0<Q_m\leq1$ there may be three positive roots for $f(r)$. We take the largest one of them to be the event horizon radius $r_e$. We find that the obtained values of $r_e$ and $r_{ph}$ are hardly distinguishable from their standard counterpart, meaning that in the case where three positive roots are present, the resulting shadow is very similar to that of a standard charged BH.

Let us now consider the extremal EEH BH, which is defined by $f(r=r_{ex})=0=f'(r=r_{ex})$, from which we obtain the following:
\begin{eqnarray}
r_{ex}^6-2r_{ex}^5+Q_m^2r_{ex}^4-\frac{2\mu Q_{m-ex}^4}{5}=0\,, \quad  5r_{ex}^5-10r_{ex}^4+4Q_{m-ex}^2r_{ex}^3=0\,.
\end{eqnarray}
These two equations are solved by the following:
\begin{eqnarray}
&&Q_{m-ex}^2=\dfrac{3 r_{ex}^4 \pm r_{ex}^{5/2} \sqrt{9 r_{ex}^3 - 48 \mu + 24 r_{ex} \mu}}{4\mu}\,, \nonumber \\
&&10 r_{ex}^4 - 6 r_{ex}^5-\dfrac{3 r_{ex}^7 \pm r_{ex}^{11/2} \sqrt{9 r_{ex}^3 - 48 \mu + 24 r_{ex} \mu}}{\mu}=0\,.
\label{ex}
\end{eqnarray}
Given different values of $\mu$, we can extract the extremal horizon radius $r_{ex}$ as well as relevant magnetic charge $Q_m$. We find that in Eq.~(\ref{ex}) the expression with positive sign is not physical as it returns an imaginary solution. However, for the case where we choose the negative sign one may obtain a real solution provided that $0<\mu<0.1$, see Table~\ref{Nu3}.

\begin{table}
\small
\begin{center}
\begin{tabular}{|c|c|c|c|c||c|c|}
\hline
$\mu$&$r_{ph}(Q_m=0.5)$&$r_{e}(Q_m=0.5)$&$r_{eff}(Q_m=0.5)$&$r_{ph}(Q_m=0.9)$&$r_{e}(Q_m=0.9)$&$r_{eff}(Q_m=0.9)$ \\
\hline
$0$&$2.82288$ &$1.86603$&$0$&$2.16708$ &$1.43589$&$0$ \\  \hline
$0.1$&$2.81676$ &$1.86614$&$0.740083$&$2.24906$ &$1.44278$&$0.992925$ \\  \hline
$0.2$&$2.81053$ &$1.86626$&$0.880112$&$2.19579$ &$1.44933$&$1.18079$\\ \hline
$0.3$&$2.80419$ &$1.86638$&$0.974004$&$2.12858$ &$1.45557$&$1.30676$\\ \hline
$0.4$ & $2.79773$ &$1.86650$&$1.04664$&$2.03321$ &$1.46153$&$1.40421$ \\ \hline
$0.5$&$2.79116$ &$1.86662$&$1.10668$&$1.77378$ &$1.46724$&$1.48477$ \\ \hline
$0.6$&$2.78445$ &$1.86674$&$1.15829$&$1.36009$ &$1.47273$& $1.55401$\\ \hline
$0.7$&$2.77761$ &$1.86686$&$1.2038$&$1.36943$ &$1.47801$&$1.61507$ \\ \hline
$0.8$&$2.77063$ &$1.86698$&$1.24467$&$1.38512$ &$1.48311$& $1.66989$\\ \hline
$0.9$&$2.7635$ &$1.86709$&$1.28186$&$1.40246$ &$1.48803$&$1.7198$\\ \hline
$1$&$2.75622$ &$1.86721$&$1.31607$&$1.42$ &$1.4928$ &$1.7657$\\ \hline
\end{tabular}
\caption{Numerical solution of Eq.~(\ref{un1}) and $f(r)=0=h(r)$ for certain values of $0.1\leq\mu\leq1$. Note that $r_{ph}$ is the radial coordinate of the photon sphere, whereas $r_e$ and $r_{eff}$ characterize the radial coordinate of the event horizon, depending on which of the two is larger.}
\label{Nu}
\end{center}
\end{table}

\begin{table}
\begin{center}
\begin{tabular}{|c|c|c|c|}
\hline
$\mu$&$r_{ph}(Q_m=2)$&$r_{e}(Q_m=2)$ &$r_{eff}(Q_m=2)$\\
\hline
$0.1$&$0.781778$ &$0.673719$&$1.48017$\\  \hline
$0.2$&$0.935749$ &$0.805683$&$1.76022$\\ \hline
$0.3$&$1.03893$ &$0.893585$&$1.94801$\\ \hline
$0.4$ & $1.11848$ &$0.961003$&$2.09327$ \\ \hline
$0.5$&$1.18396$ &$1.01624$ &$2.21336$\\  \hline
$0.6$&$1.23996$ &$1.0633$ &$2.31658$\\ \hline
$0.7$&$1.2891$ &$1.10445$ &$2.4076$\\ \hline
$0.8$&$1.33301$ &$1.14109$ &$2.48933$\\ \hline
$0.9$&$1.37277$ &$1.17418$&$2.56372$\\ \hline
$1$&$1.40917$ &$1.20438$&$2.63215$\\ \hline
\end{tabular}
\caption{Numerical solution of Eq.~(\ref{un1}) and $f(r)=0=h(r)$ for certain values of $0.1\leq\mu\leq1$. }
\label{Nu2}
\end{center}
\end{table}

\begin{table}
\begin{center}
\begin{tabular}{|c|c|c|c|c|}
\hline
$\mu$&$Q_{m-ex}$&$r_{ph}$&$r_{ex}$ &$r_{eff}$\\
\hline
$0.01$&$1.0005$ &$1.98878$&$0.985453$&$0.588713$\\  \hline
$0.02$&$1.01461$ &$1.91462$&$0.967641$&$0.699827$\\ \hline
$0.03$&$1.02323$ &$1.85499$&$0.94419$&$0.773713$\\ \hline
$0.04$ & $1.03343$ &$1.76873$&$0.907418$&$0.829188$ \\ \hline
\end{tabular}
\caption{Numerical solution of Eqs.~(\ref{ex},\ref{un1}) for certain values of $0\leq\mu<0.1$. The extremal magnetic charge and horizon radius in the standard RN case are given by $Q_{m-ex}=1$ and $r_{ex}=2$ respectively.}
\label{Nu3}
\end{center}
\end{table}

\begin{figure}[!ht]
	\begin{center}
\includegraphics[scale=0.45]{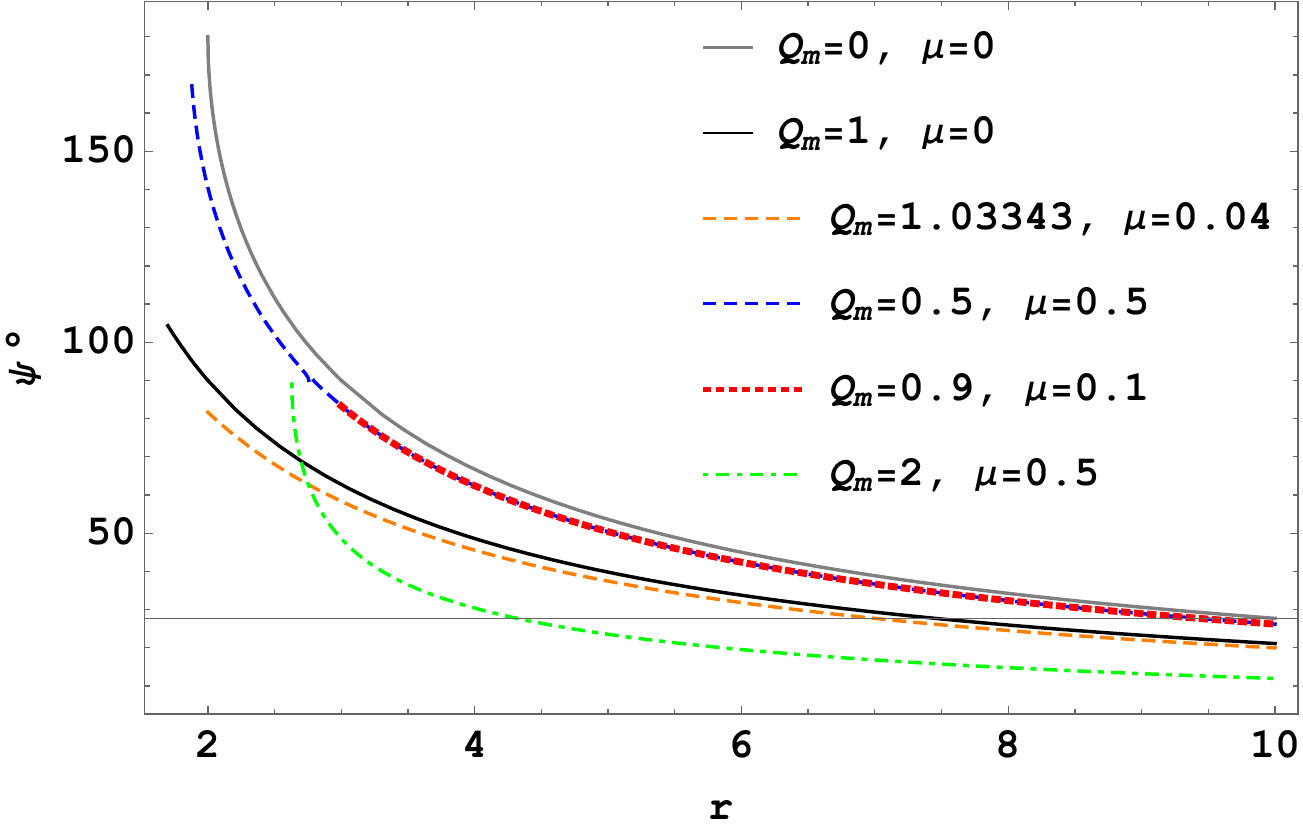}
\caption{Plot of the angle between the four-vector tangent to the path of a photon from an unstable circular orbit and the position coordinate of a static distant observer, $\psi$, versus radial coordinate for several values of $Q_m$ and $\mu$ in the Einstein-Euler-Heisenberg model.}
\label{Mag1}
\end{center}
\end{figure}

Finally, let us consider the four-vector $K^{\mu}$ tangent to the photon's path so that using Eqs. (\ref{E}-\ref{V}), this is given by:
\begin{eqnarray}
K^{\mu}=\frac{dx^{\mu}}{d\tau}=\left(\frac{r^2 h(r)}{b f(r)g(r)} ,\sqrt{V(r)},0,1\right)~.
\end{eqnarray}
Choosing the position coordinate of a static distant observer located at $r=r_o$, as $D^{\mu}=\left(0,r,0,0 \right)$, the angle between $K^{\mu}$ and $D^{\mu}$, takes the following form:
\begin{eqnarray}\label{psi}
\psi=\cos^{-1} \left ( \sqrt{\dfrac{ g(r) V(r)}{g(r) V(r)+f(r) h(r)r^2}} \right ) \,.
\end{eqnarray}
By inserting our numerical results from Tables~\ref{Nu},\ref{Nu2} into the above relation, we can track the non-linearity effect arising from EEH electrodynamics on the shadow shape. As shown in Fig.~\ref{Mag1} for a few representative chosen values of the magnetic charge $Q_m$ and EEH coupling $\mu$, the angle between the position coordinate of a static distant observer and the four-vector tangent to the photon's path is smaller than Schwarzschild case. However, if we compare this angle to that of the standard extremal charged RN BH, we see that the angle becomes larger for $Q_m \leq 1$ and smaller for $Q_m > 1$. This means that the angular size of the magnetically charged BH in EEH gravity can be larger or smaller than that of the corresponding extremal RN BH. Note that what we mean by the size of the shadow is indeed the shadow angular size.
\begin{figure}[!ht]
	\begin{center}
\includegraphics[scale=0.45]{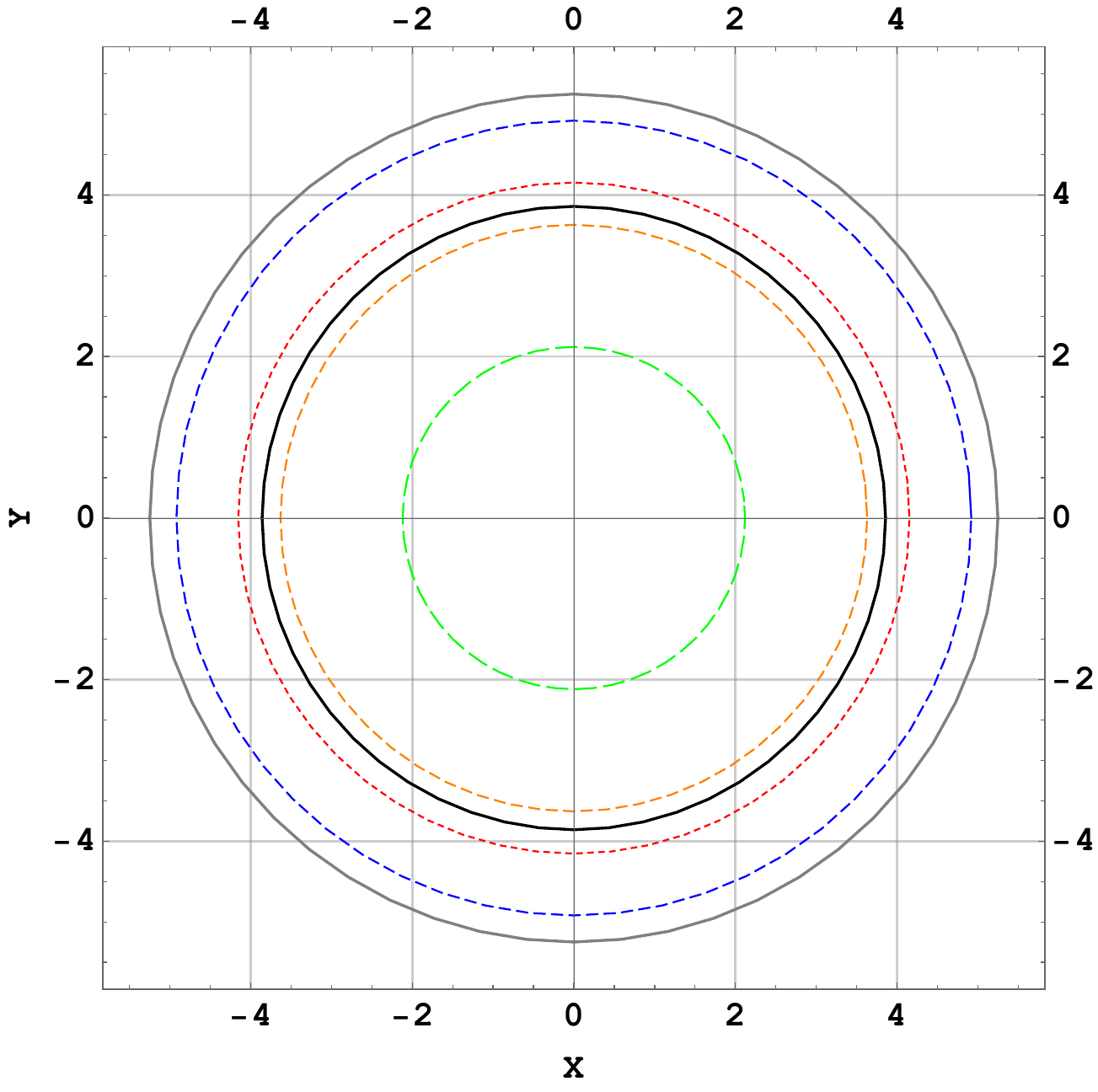}
\caption{Shadows of Einstein-Euler-Heisenberg non-linear BHs, as viewed by a distant observer, with the same colour-coding as in Fig.~\ref{Mag1}. Note that both axes are in units of BH mass $M$.}
\label{Mag2}
\end{center}
\end{figure}

In Fig.~\ref{Mag2} we plot the shadow resulting from the chosen values of $Q_m$ and $\mu$ used in Fig.~\ref{Mag1}. We clearly see that as $Q_m$ is increased above $1$, the shadow angular size shrinks with respect to the standard extremal RN case with $Q_e=1$. similarly, as $Q_m$ is decreased below $1$, the angular size increases. Finally, as $Q_m \to 0$, the shadow size approaches the Schwarzschild limit, as expected.

We can therefore expect the shadow of M87* detected by the Event Horizon Telescope to be able to set limits on $Q_m$, given that the anguar size of the observed shadow is consistent with that of a Schwarzschild BH. In particular, we expect to get an upper limit on $Q_m$ since, if $Q_m$ is increased too much, the shadow becomes significantly smaller than the standard Schwarzschild shadow. On the other hand, we see that the EEH coupling $\mu$ has a very limited effect on the angular size of the shadow, and therefore we do not expect it to be subject to tight constraints from the shadow of M87*. We will address these issues in Sec.~\ref{sec:ehtshadow}.

\section{Shadows regular magnetic Einstein-Bronnikov black holes}
\label{sec:shadowregular}

In this section we will repeat the calculations of the previous section in the case of Einstein-Bronnikov gravity, a particular non-linear electrodynamics theory which only makes use of the relativistic invariant $U$ and not of $W$, and wherein one can obtain regular BHs.

\subsection{Static regular black hole}
\label{subsec:staticregular}

Let us consider the following action for Einstein-Bronnikov (EB) gravity~\cite{Bronnikov:2000vy}:
\begin{eqnarray}
S=\frac{1}{16\pi}\int d^4x\,,\sqrt{-g} \left [ R-\mathcal{L}(U) \right ]\,,
\label{Ln}
\end{eqnarray}
where:
\begin{eqnarray}
\mathcal{L}(U)=F_{\mu\nu}F^{\mu\nu}\cosh^{-2}\left[ a \left( F_{\alpha\beta}F^{\alpha\beta}/2\right)^{1/4}\right]\,.
\end{eqnarray}
The existence of the constant parameter $a$ removes the singularity at the centre of the BH. When considering $a\rightarrow 0$, the standard Einstein-Maxwell Lagrangian is recovered. Within the EB NLED theory, the gravitational field equations are given by:
\begin{eqnarray}
G_{\mu\nu}=8\pi T_{\mu\nu}=\left(2\mathcal{L}_{U}\, F_{\rho \mu}F^{\rho}_{\nu}-\frac{1}{2}g_{\mu\nu} {\cal L}\left( U\right) \right)\,,
\label{EE}
\end{eqnarray}
and:
\begin{eqnarray}\label{maxwell2}
\nabla _{\mu}\left(\mathcal{L}_U F^{\mu\nu}\right) =0~,~~~~\nabla_{\mu}\,^{\star }F^{\mu\nu}=0\,.
\end{eqnarray}

We again choose to study a purely magnetic BH, taking the gauge field to be given by:
\begin{eqnarray}
A_\mu=-Q_m\cos\theta \delta_{\mu}^{\varphi}\,.
\label{gaugeR}
\end{eqnarray}
Taking again a SSS ansatz, the relevant metric function analogous to Eq.~(\ref{ds}) is now given by the following:
\begin{eqnarray}
f_{\rm EB}(r)=1-\frac{Q_m^{3/2}}{a r }\left(1-\tanh\frac{a Q_m^{1/2}}{r} \right)
\end{eqnarray}
Here the constant parameter $a$ is linked to the black hole mass and magnetic charge via $a=Q_m^{3/2}/2M$. Therefore, the metric function takes the following form:
\begin{eqnarray}
f_{\rm EB}(r)=1-\frac{2M}{r} \left ( 1-\tanh\frac{Q_m^2}{2Mr} \right ) \,.
\label{laps}
\end{eqnarray}
It is straightforward to see that for $Q_m=0$ or small values of $Q_m$ (weak field limit), the metric function Eq.~(\ref{laps}) reduces to the standard Schwarzschild or RN metric function respectively. An attractive feature of this metric function is that one can easily see that $f_{\rm EB}(r)$ is regular as it approaches $1$ as $r \to 0$ unlike what occurs for a standard RN BH.

\subsection{Effective geometry induced by non-linear Bronnikov electrodynamics and resulting black hole shadows}
\label{subsec:effectivegeometry2}

\begin{figure}[!ht]
	\begin{center}
\includegraphics[scale=0.45]{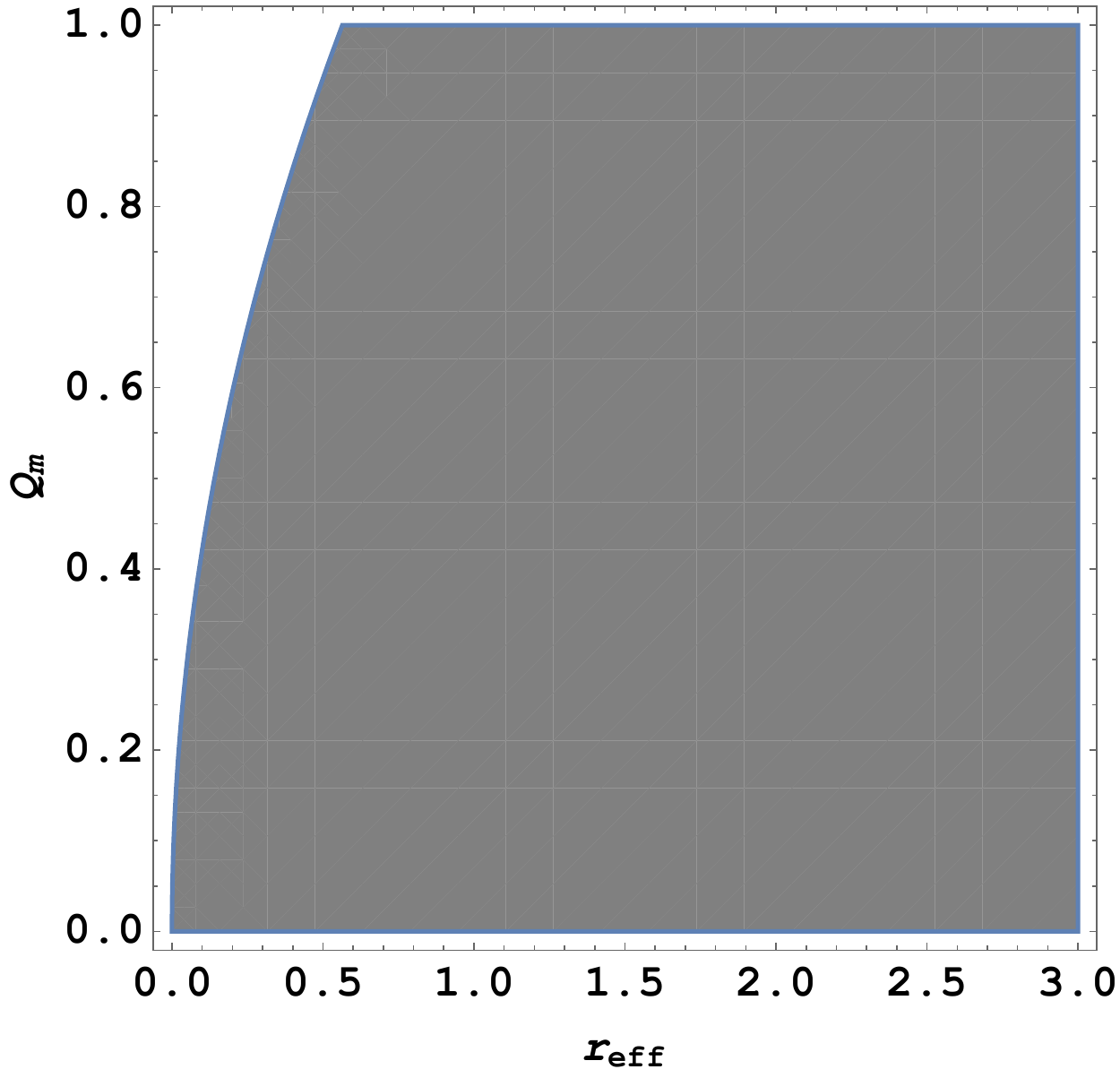}
\caption{The allowed region in the $(r_{eff}, Q_m)$ parameter space given by the positive definiteness of $g_{\rm EB}(r)$ and $h_{\rm EB}(r)$ is given by the grey shaded area.}
\label{Mag3}
\end{center}
\end{figure}

By using the expression Eq.~(\ref{geral1}) related to Lagrangian with one relativistic invariant, the effective geometry for light rays on the Einstein-Bronnikov background is given by:
\begin{eqnarray}\label{effB}
ds^2_{\rm EB}=g_{\rm EB}(r)\left( -f_{\rm EB}(r)dt^2+\frac{dr^2}{f_{\rm EB}(r)}\right)+h_{\rm EB}(r)\left (r^2 d\theta^2+r^2\sin^2\theta d\phi^2\right)\;,
\end{eqnarray}
with the following effective metric functions:
\begin{eqnarray}
&g_{\rm EB}(r)=\frac{2-\frac{Qm^2}{2Mr}\tanh\frac{Qm^2}{2Mr}}{2\cosh^2\frac{Qm^2}{2Mr} }\,,\\
&h_{\rm EB}(r)=\frac{4-\frac{7Qm^2}{2Mr}\tanh\frac{Qm^2}{2Mr}-\frac{3Qm^4}{(2Mr)^2}\cosh^{-1}\frac{Qm^2}{2Mr}+\frac{Qm^4}{2(Mr)^2}}{4\cosh^2\frac{Qm^2}{2Mr} }\,,
\end{eqnarray}
As we see from Eq.~(\ref{effB}), the effective geometry of the Bronnikov spacetime is again spherically symmetric and static as we expected. We again work in units of mass, setting $M=1$. The condition of positive definiteness for the effective metric functions $g_{\rm EB}(r)$ and $h_{\rm EB}(r)$ allows us to identify to find the allowed region of parameter space in the $(r_{eff},Q_m)$ plane, as illustrated in Fig.~\ref{Mag3} in the grey shaded area.
\begin{figure}[!ht]
	\begin{center}
\includegraphics[scale=0.45]{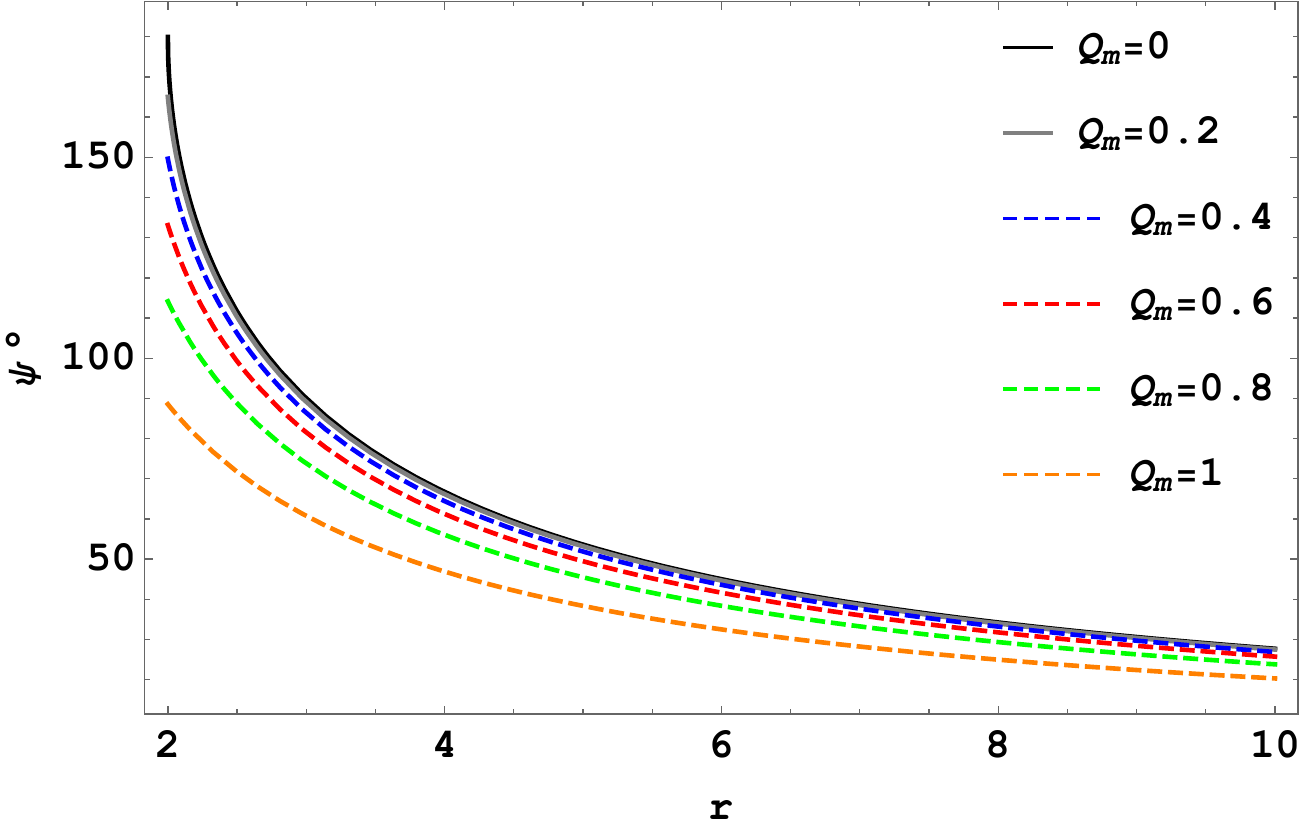}
\caption{Plot of the angle between the four-vector tangent to the path of a photon from an unstable circular orbit and the position coordinate of a static distant observer, $\psi$, versus radial coordinate for several values of $Q_m$ and $\mu$ in the Einstein-Bronnikov model.}
\label{Mag4}
\end{center}
\end{figure}

As for the non-regular BHs we studied in the previous section, we can repeat exactly the steps between Eqs.~(\ref{eff1}-\ref{un}) to obtain the following expressions describing unstable circular orbits:
\begin{eqnarray}\label{74}
b^{-2}=\frac{4 \left(2 \tanh \frac{Q_m^2}{2 r}+r-2\right) \left ( 4 r-Q_m^2 \tanh \frac{Q_m^2}{2r}\right)}{r^2 \left(2 \left(Q_m^4+8r^2\right)-Q_m^2 \left(7 r \sinh\frac{Q_m^2}{r}+3 Q_m^2 \right ) \text{sech}^2\frac{Q_m^2}{2 r} \right ) }\,,
\end{eqnarray}
and:
$$ \bigg(Q_m^8 (6-3r)-28Q_m^6r\bigg)
   \text{sech}\frac{Q_m^2}{2 r}+\bigg(Q_m^6
   (4r-r^2)+50 Q_m^4 r^2+4 Q_m^2 (44r^3-13r^4) +192 r^4\bigg)
   \sinh \frac{3 Q_m^2}{2 r}-$$
   $$ \bigg(4Q_m^6 r^2+Q_m^4 (50r^2-14 r^3)+176Q_m^2 r^3-64(r^5-3r^4)
   \bigg) \cosh\frac{3 Q_m^2}{2 r}+$$
   $$ \bigg(4 Q_m^8+Q_m^6 (28r-19 r^2)-230 Q_m^4 r^2+4
   Q_m^2 (44r^3-13r^4) +192 r^4\bigg) \sinh
   \frac{Q_m^2}{2 r}+$$
   \begin{eqnarray}\label{75}
   \bigg(4Q_m^8 (r-2)+24
   Q_m^6 r+Q_m^4(26 r^3-14r^2)+304Q_m^2 r^3+192(r^5-3r^4)
   \bigg) \cosh \frac{Q_m^2}{2 r}=0\,. \nonumber \\
   \end{eqnarray}
\begin{table}
	\begin{center}
		\begin{tabular}{|c|c|c|c|c|}
			\hline
			$Q_m$&$r_{ph}(EB)$&$r_{e}(EB)$&$r_{ph}(RN)$&$r_{e}(RN)$\\
			\hline
			$0.1$&$2.99331$ &$1.99499$&$2.99332$&$1.99499$\\  \hline
			$0.2$&$2.97303$ &$1.9798$&$2.97309$&$1.9798$\\ \hline
			$0.3$&$2.93841$ &$1.95395$&$2.93875$&$1.95394$\\ \hline
			$0.4$ & $2.88816$ &$1.91657$&$2.88924$ &$1.91652$\\ \hline
			$0.5$&$2.8202$ &$1.86624$ &$2.82288$&$1.86603$\\  \hline
			$0.6$&$2.73132$ &$1.80075$ &$2.73693$&$1.80$\\ \hline
			$0.7$&$2.61643$ &$1.71645$&$2.62694$&$1.71414$\\ \hline
			$0.8$&$2.46685$ &$1.6069$ &$2.48489$&$1.60$\\ \hline
			$0.9$&$2.27028$ &$1.45847$&$2.29373$&$1.43589$ \\ \hline
			$1$&$1.96682$ &$1.22771$&$2$&$1$\\ \hline
		\end{tabular}
		\caption{Numerical solution of Eqs. (\ref{75}) and $f_{\rm EB}(r)=0$ for certain values of magnetic charge $Q_m$, given in terms of the radial coordinate of the photon sphere $r_{ph}$ and of the event horizon $r_e$.}
		\label{Nu4}
	\end{center}
\end{table}
\begin{figure}[!ht]
	\begin{center}
		\includegraphics[scale=0.45]{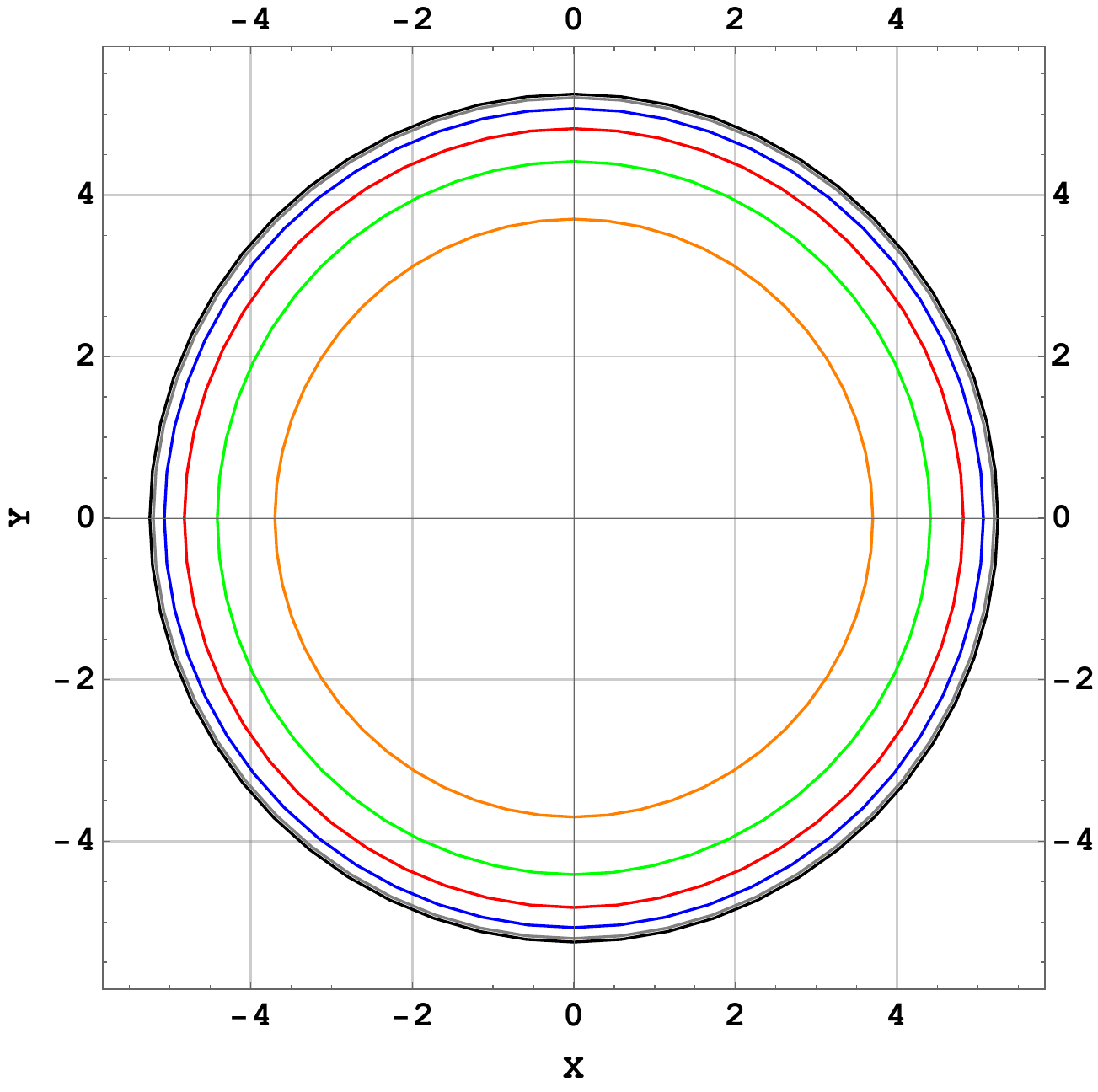}
		\caption{Shadows of Einstein-Bronnikov non-linear BHs, as viewed by a distant observer. The colour-coding corresponds to different values of $Q_m$: $Q_m=0,0.2,0.4,0.6,0.8,1$ moving from the outermost black curve toward the innermost orange curve, respectively. Note that both axes are in units of BH mass $M$.}
		\label{Mag5}
	\end{center}
\end{figure}

The numerical solution to the above involved expression returns us the radial coordinates of the unstable circular orbits for various values of the magnetic charge $0<Q_m\leq1$, see Table~\ref{Nu4}. Unlike the Euler-Heisenberg case, we find that the photon's geodesic can cover the entire region outside the event horizon since $r_{eff}<r_e$. Using Eq.~(\ref{psi}), we show in Fig.~\ref{Mag4} the angle between the position coordinate of a static distant observer and the four-vector tangent to the photon's path in the Einstein-Bronnikov model. To have a better intuitive understanding of the effect of the magnetic charge on the shadow size in Einstein-Bronnikov spacetime, in Fig.~\ref{Mag5} we plot the resulting BH shadows obtained for the same values of $Q_m$ considered in Fig.~\ref{Mag4}. Both figures explicitly show that in the Einstein- Bronnikov spacetime the shadow size is smaller than its standard Schwarzschild or RN counterparts.
\begin{figure}[!ht]
	\begin{center}
		\includegraphics[scale=0.5]{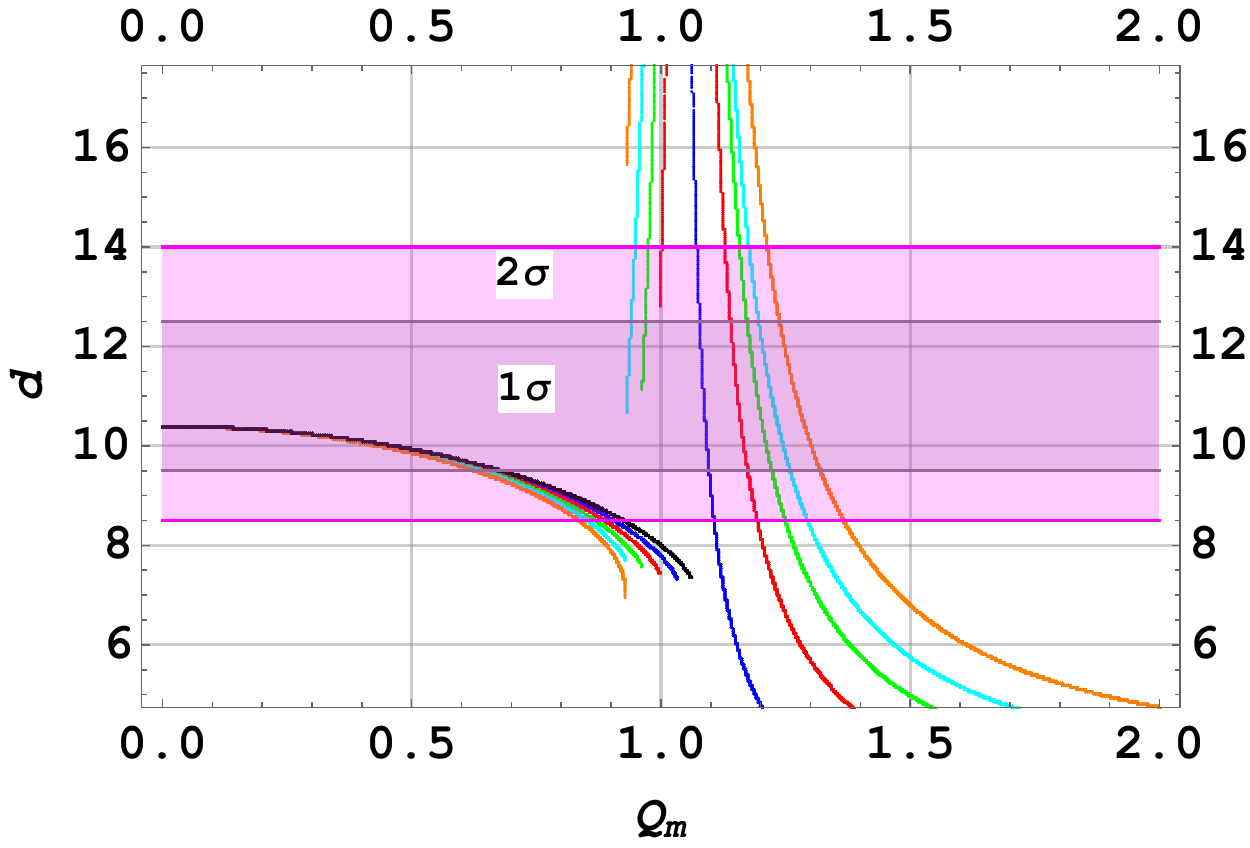}
		\caption{Diameter of the shadow of Einstein-Euler-Heisenberg black holes $d$ as a function of the magnetic charge $Q_m$, for various values of the non-linear coupling strength $\mu$: $\mu=0$ (black), $0.1$ (dark blue), $0.2$ (red), $0.3$ (green), $0.4$ (light blue), and $0.5$ (orange). The shaded regions indicate the values of $d$ consistent with the shadow of the supermassive BH M87* detected by the Event Horizon Telescope, see Eq.~(\ref{size}). The grey shaded region gives the $1\sigma$ confidence region for $d$, whereas the magenta shaded region gives the $2\sigma$ confidence region. See the main text for further discussions on the behaviour of the curves in the figure, in particular concerning the apparent divergence in the shadow size for $Q_m \simeq 1$.}
		\label{Mag6}
	\end{center}
\end{figure}
\begin{figure}[!ht]
	\begin{center}
		\includegraphics[scale=0.5]{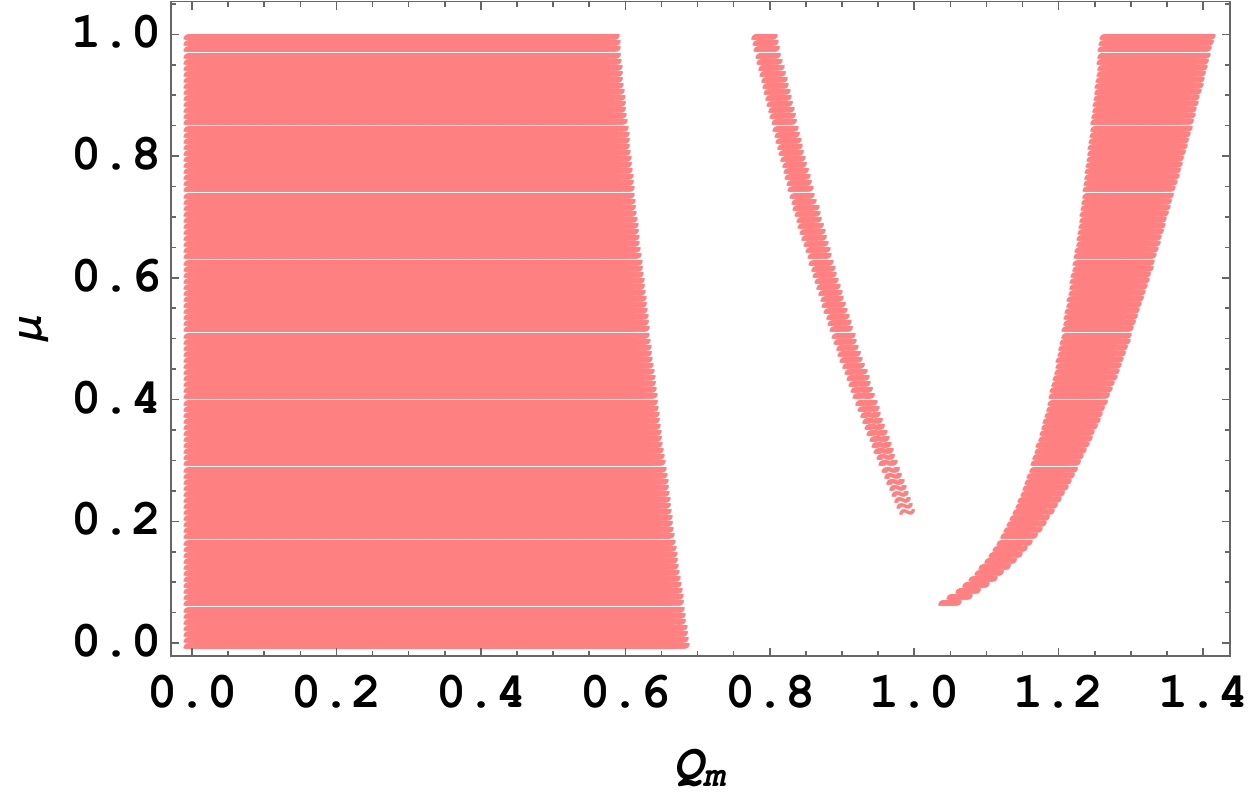}
		\includegraphics[scale=0.5]{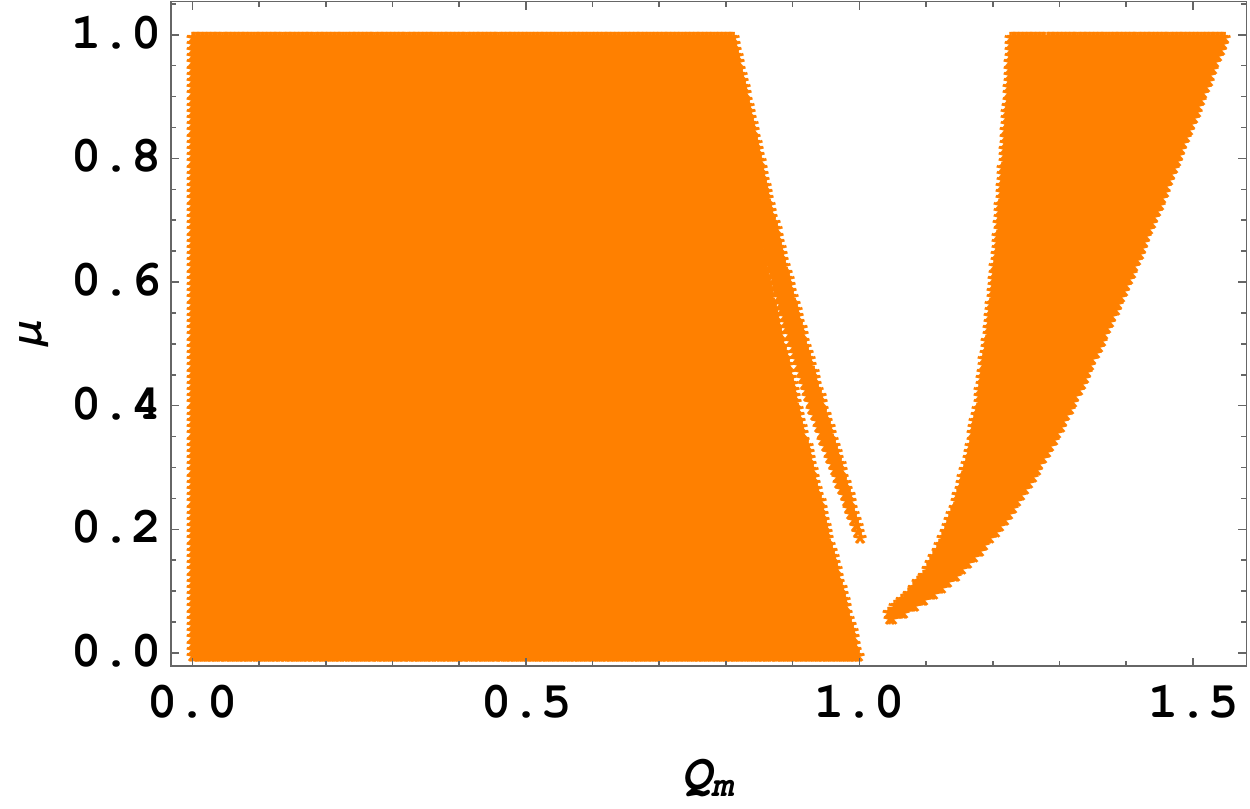}
		\caption{Allowed region in the parameter space of $Q_m$ (magnetic charge) and $\mu$ (nonlinear coupling strength) for Einstein-Euler-Heisenberg BHs, given the shadow of the supermassive BH M87* detected by the Event Horizon Collaboration. In the upper panel the region shaded in pink is consistent with the $1\sigma$ confidence region for the diameter of the shadow of M87* as given by Eq.~(\ref{size}). Similarly, the orange shaded region in the lower panel is consistent with the $2\sigma$ confidence region for the diameter of M87*.}
		\label{Mag7}
	\end{center}
\end{figure}
\section{Comparison with the Event Horizon Telescope's shadow of M87*}
\label{sec:ehtshadow}

In this section, we compare the shadows for BHs within the Einstein-Euler-Heisenberg and Einstein-Bronnikov models obtained in Sec.~\ref{sec:shadownonregular} and Sec.~\ref{sec:shadowregular} with the shadow of M87* detected by the Event Horizon Telescope~\cite{Akiyama:2019cqa}. As we saw earlier in Fig.~\ref{Mag2} and Fig.~\ref{Mag5}, the shadow size depends rather strongly on the value of the magnetic charge $Q_m$. Therefore, it should at least in principle be possible to constrain this quantity using the EHT observation.

As reported in~\cite{Akiyama:2019cqa}, the angular size of the shadow of M87* as detected by the EHT is $\delta = (42 \pm 3)\,\mu{\rm as}$, whereas following~\cite{Akiyama:2019eap} the distance to M87* is $D = 16.8^{+0.8}_{-0.7}\,{\rm Mpc}$ and the mass of M87* is $M = (6.5 \pm 0.9) \times 10^9\,M_{\odot}$. Combining this information as in~\cite{Bambi:2019tjh} we can infer the diameter of the shadow in units of mass $d_{M87*}$ to be:
\begin{eqnarray}
d_{M87*} \equiv \frac{D\delta}{M} \approx 11.0 \pm 1.5\,.
\label{size}
\end{eqnarray}

\begin{figure}[!ht]
	\begin{center}
		\includegraphics[scale=0.5]{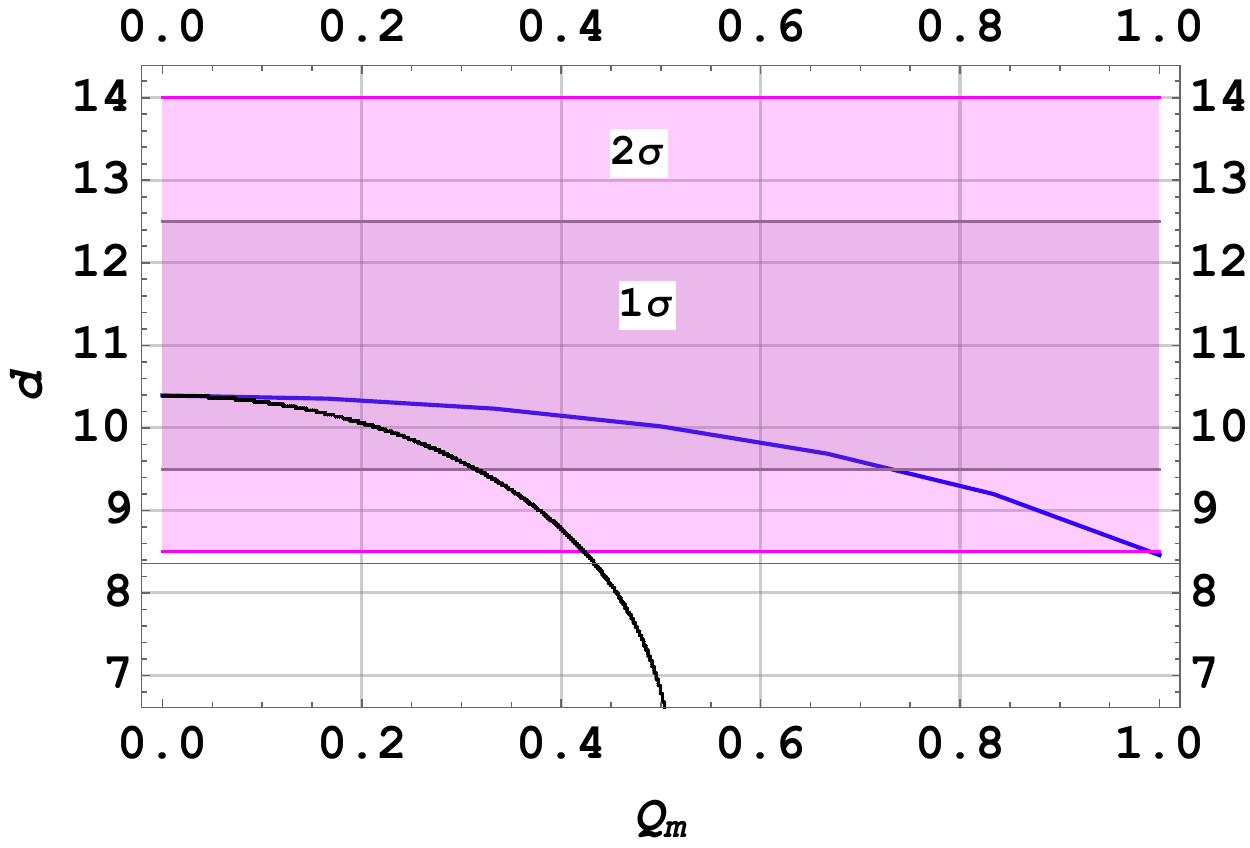}
		\caption{The blue curve plots the diameter of the shadow of Einstein-Bronnikov black holes as a function of magnetic charge $Q_m$. For comparison, the black curve plots the same quantity for a Reissner-Nordstr\"{o}m BH (in this case the $x$ axis gives the RN BH electric charge). The shaded regions indicate the values of $d$ consistent with the shadow of the supermassive BH M87* detected by the Event Horizon Telescope, see Eq.~(\ref{size}). The grey shaded region gives the $1\sigma$ confidence region for $d$, whereas the magenta shaded region gives the $2\sigma$ confidence region.}
		\label{Mag8}
	\end{center}
\end{figure}

The detected diameter of M87*'s shadow, as given in Eq.~(\ref{size}), is remarkably consistent with that of the Schwarzschild BH, as we can see from Figs.~\ref{Mag2} and~\ref{Mag5}. Within $1\sigma$ uncertainties, we see that $9.5 \lesssim d_{M87*} \lesssim 12.5$, whereas within $2\sigma$ uncertainties $8.0 \lesssim d_{M87*} \lesssim 14.0$. The upper limits of these intervals are mostly irrelevant for the purpose of our discussion because as we have seen in Figs.~\ref{Mag2} and~\ref{Mag5}, increasing the magnetic charge $Q_m$ (and in the case of the EEH model also the EEH non-linear electrodynamics coupling $\mu$, albeit the latter has a marginal effect on the shadow) always leads to a smaller angular size for a shadow, and never a larger one. From these considerations we can expect to set an upper limit on $Q_m$ both in the EEH and in the EB model, for if $Q_m$ becomes too large then the diameter of the shadow would become too small and inconsistent with the detection of the Event Horizon Telescope.

We perform a parameter scan of the $Q_m$-$\mu$ parameter space for the magnetically charged Einstein-Euler-Heisenberg BH, and of the $Q_m$ parameter space for the Einstein-Bronnikov BH. For each point in the parameter space, we compute the diameter of the resulting shadow. We then check for what regions of parameter space the EHT constraint in Eq.~(\ref{size}) is satisfied, considering both $1\sigma$ and $2\sigma$ confidence regions as discussed earlier. For the EEH model we consider values of $\mu<1$, as for larger values we lose perturbative control over the theory.

We begin by considering the Einstein-Euler-Heisenberg BH studied in Sec.~\ref{sec:shadownonregular}. In Fig.~\ref{Mag6} we plot the diameter of the resulting BH shadow as a function of the magnetic charge $Q_m$, for various values of the NLED coupling strength $\mu$, together with $1\sigma$ and $2\sigma$ confidence intervals on the diameter of the shadow of M87* as reported in Eq.~(\ref{size}). As we see from the figure and as expected earlier, we can certainly set an upper limit on $Q_m$ for if this quantity increases too much, the size of the shadow becomes too small and inconsistent with observations. From the figure we also confirm our earlier finding that the effect of $\mu$ on the shadow size is rather limited.

Fig.~\ref{Mag6} deserves a further comment. As we see, around $Q_m \approx 1$, the sizes of the shadows appear to diverge. To understand this, we can recall from Eq.~(\ref{b}) that the size of the shadow scales as $\sqrt{h/(f \times g)}$. For small $Q_m$ we have that $h>0$ and $g>0$, and a shadow of finite size. As $Q_m$ is increased, $h \to 0$ while $f$ and $g$ remain positive, so the shadow decreases in size. As we keep increasing $Q_m$, we first hit a region excluded by the spacetime geometry (see earlier discussion). After that we find $f \approx h \approx 0$ and $g>0$, with the shadow size increasing. As $g \to 0$ as well, the shadow size appears to diverge. After that both $g$ and $h$ switch sign, in such a way that the combination $h/(f \times g)$ remains positive and the shadow is again finite in size.

In Fig.~\ref{Mag7} we shade the region of $Q_m$-$\mu$ parameter space where the resulting BH shadow has a size compatible with the EHT detection ($1\sigma$ confidence interval in the upper panel, $2\sigma$ confidence interval in the lower panel). We see from these figures that we are essentially unable to set any meaningful limit on $\mu$ as expected, whereas we can set a rough upper limit of $Q_m<1.5M$ (recall that we had worked in units of $M=1$). This limit is to some extent affected by our choice of restricting the coupling strength to $\mu<1$, in order to maintain perturbative control on the theory. Nonetheless, given the fact that the influence of $\mu$ on the shadow is quite limited, we can in all generality recast our observational limit as $Q_m \lesssim {\cal O}(M)$.

It is interesting to note that, by increasing the value of $\mu$, there is the possibility of considering values of the magnetic charge beyond the extremal limit ($Q_m>M=1$) while maintaining consistency with the EHT observations. As a consistency check one can see explicitly in Figs.~\ref{Mag6} and~\ref{Mag7} that for case of $\mu=0$ the region $Q_m>1$ is not allowed.

We then repeat the same scan for the EB model, focusing only on the magnetic charge $Q_m$. From Fig.~\ref{Mag8} we see that, for a given value of the electric (respectively magnetic) charge, a RN BH (black curve) will have a smaller shadow than the corresponding EB BH (blue curve). From the parameter scan, we find a rough upper limit of $Q_m<0.7M$ at $1\sigma$ and $Q_m<1.0M$ at $2\sigma$. Again as in the EEH case, we therefore again find a limit $Q_m \lesssim {\cal O}(M)$.

An important caveat concerning our comparison to the shadow of M87* detected by the EHT collaboration is in order. In our work we have considered static (non-rotating) solutions. However, it is likely that M87* (as most BHs) is rotating, and the rotation could potentially have an important effect on the shadow. It is known that the effect of introducing angular momentum is that of making the shadow slightly asymmetric (in particular the shadow flattens on the side corresponding to photons with angular momentum aligned with the BH spin, as the effective potential is shallower in that case) and hence less circular, see e.g.~\cite{Bardeen:1973tla,Bambi:2019tjh}. This deviation is, however, very small and only important at high observation angles (see e.g. Fig.~1 in~\cite{Bambi:2019tjh}). For M87* there is very good reason to believe that the mechanism powering the jet is closely related to the Blandford-Znajek mechanism~\cite{Akiyama:2019fyp,Blandford:1977ds}. Under this assumption, the observation angle of M87* (the angle between the BH angular momentum and the line-of-sight) is very close to the jet angle, $\theta \approx 17^{\circ}$. Such a low observation angle makes the effect of rotation even less important, as the latter are important only when $\theta \to \pi/2$, or equivalently when the BH is viewed edge-on. Nonetheless, it would be worth further examining this conclusion, and possibly explicitly construct rotating solutions for the non-linear magnetically charged EEH and EB BHs we have studied in this work, for instance by adopting the Newman-Janis algorithm. We leave this issue for future work.

\section{Conclusions}
\label{sec:conclusions}

Non-linear electrodynamics (NLED) theories constitute well-motivated extensions to QED in the strong-field regime. Perhaps more interestingly, there is mounting evidence that NLED theories might harbor regular black hole (BH) solutions, which thus address the thorny issue of singularities from continuous gravitational collapse in GR. In this work, we have considered two well-known NLED theories coupled to GR: the Einstein-Euler-Heisenberg (EEH) theory and the Einstein-Bronnikov (EB) theory. We have first of all considered solutions for magnetically charged BHs within both theories. It is known that NLED results in the motion of photons being along geodesics of a so-called effective geometry, an effect which only recently has been appreciated in the literature. Taking this effect carefully into account, we have determined the shadows of magnetically charged BHs within the EEH and EB theories, as a function of the magnetic charge $Q_m$.

We have then confronted the resulting shadows with the observed the shadow of the supermassive BH M87*, recently detected by the Event Horizon Telescope collaboration. Using the inferred diameter of M87*'s shadow, which is highly consistent with that of a Schwarzschild BH, we have set a rough upper limit of $Q_m \lesssim {\cal O}(M)$ on the magnetic charge, where $M$ is the BH mass. On the other hand, the NLED coupling strength for the EEH theory, $\mu$, remains basically unconstrained within the region $\mu<1$.

Overall, our results present the first astrophysical constraints on NLED. Moreover, ours is among the first works to provide constraints on new physics beyond the Standard Model from the Event Horizon Telescope detection of the shadow of M87*. We leave further extensions of our results, such as a study of the effect of rotation on the shadows of magnetically charged BHs, to future work.

\acknowledgments

A.A. and M.Kh. appreciate H.Firouzjahi for his supports. S.V. acknowledges support from the Isaac Newton Trust and the Kavli Foundation through a Newton-Kavli Fellowship, and acknowledges a College Research Associateship at Homerton College, University of Cambridge. D.F.M. acknowledges support from the Research Council of Norway.

\bibliographystyle{JHEP}
\bibliography{BH}

\providecommand{\href}[2]{#2}\begingroup\raggedright\begin{thebibliography}{100}

\bibitem{Einstein:1916vd}
A.~Einstein, \emph{{The Foundation of the General Theory of Relativity}},
  \href{https://doi.org/10.1002/andp.200590044,
  10.1002/andp.19163540702}{\emph{Annalen Phys.} {\bfseries 49} (1916)
  769--822}.

\bibitem{Schwarzschild:1916uq}
K.~Schwarzschild, \emph{{On the gravitational field of a mass point according
  to Einstein's theory}}, {\emph{Sitzungsber. Preuss. Akad. Wiss. Berlin (Math.
  Phys.)} {\bfseries 1916} (1916) 189--196},
  [\href{https://arxiv.org/abs/physics/9905030}{{\ttfamily physics/9905030}}].

\bibitem{Penrose:1964wq}
R.~Penrose, \emph{{Gravitational collapse and space-time singularities}},
  \href{https://doi.org/10.1103/PhysRevLett.14.57}{\emph{Phys. Rev. Lett.}
  {\bfseries 14} (1965) 57--59}.

\bibitem{Hawking:1976ra}
S.~W. Hawking, \emph{{Breakdown of Predictability in Gravitational Collapse}},
  \href{https://doi.org/10.1103/PhysRevD.14.2460}{\emph{Phys. Rev.} {\bfseries
  D14} (1976) 2460--2473}.

\bibitem{Giddings:2017jts}
S.~B. Giddings, \emph{{Astronomical tests for quantum black hole structure}},
  \href{https://doi.org/10.1038/s41550-017-0067}{\emph{Nat. Astron.} {\bfseries
  1} (2017) 0067}, [\href{https://arxiv.org/abs/1703.03387}{{\ttfamily
  1703.03387}}].

\bibitem{Giddings:2019jwy}
S.~B. Giddings, \emph{{Searching for quantum black hole structure with the
  Event Horizon Telescope}},
  \href{https://doi.org/10.3390/universe5090201}{\emph{Universe} {\bfseries 5}
  (2019) 201}, [\href{https://arxiv.org/abs/1904.05287}{{\ttfamily
  1904.05287}}].

\bibitem{LyndenBell:1969yx}
D.~Lynden-Bell, \emph{{Galactic nuclei as collapsed old quasars}},
  \href{https://doi.org/10.1038/223690a0}{\emph{Nature} {\bfseries 223} (1969)
  690}.

\bibitem{Kormendy:1995er}
J.~Kormendy and D.~Richstone, \emph{{Inward bound: The Search for supermassive
  black holes in galactic nuclei}},
  \href{https://doi.org/10.1146/annurev.aa.33.090195.003053}{\emph{Ann. Rev.
  Astron. Astrophys.} {\bfseries 33} (1995) 581}.

\bibitem{Bambi:2019xzp}
C.~Bambi, \emph{{Astrophysical Black Holes: A Review}},  2019,
  \href{https://arxiv.org/abs/1906.03871}{{\ttfamily 1906.03871}}.

\bibitem{Luminet:1979nyg}
J.~P. Luminet, \emph{{Image of a spherical black hole with thin accretion
  disk}}, {\emph{Astron. Astrophys.} {\bfseries 75} (1979) 228--235}.

\bibitem{Lu:2014zja}
R.-S. Lu, A.~E. Broderick, F.~Baron, J.~D. Monnier, V.~L. Fish, S.~S. Doeleman
  et~al., \emph{{Imaging the Supermassive Black Hole Shadow and Jet Base of M87
  with the Event Horizon Telescope}},
  \href{https://doi.org/10.1088/0004-637X/788/2/120}{\emph{Astrophys. J.}
  {\bfseries 788} (2014) 120},
  [\href{https://arxiv.org/abs/1404.7095}{{\ttfamily 1404.7095}}].

\bibitem{Cunha:2018acu}
P.~V.~P. Cunha and C.~A.~R. Herdeiro, \emph{{Shadows and strong gravitational
  lensing: a brief review}},
  \href{https://doi.org/10.1007/s10714-018-2361-9}{\emph{Gen. Rel. Grav.}
  {\bfseries 50} (2018) 42},
  [\href{https://arxiv.org/abs/1801.00860}{{\ttfamily 1801.00860}}].

\bibitem{Gralla:2019xty}
S.~E. Gralla, D.~E. Holz and R.~M. Wald, \emph{{Black Hole Shadows, Photon
  Rings, and Lensing Rings}},
  \href{https://doi.org/10.1103/PhysRevD.100.024018}{\emph{Phys. Rev.}
  {\bfseries D100} (2019) 024018},
  [\href{https://arxiv.org/abs/1906.00873}{{\ttfamily 1906.00873}}].

\bibitem{Narayan:2019imo}
R.~Narayan, M.~D. Johnson and C.~F. Gammie, \emph{{The Shadow of a Spherically
  Accreting Black Hole}},
  \href{https://doi.org/10.3847/2041-8213/ab518c}{\emph{Astrophys. J.}
  {\bfseries 885} (2019) L33},
  [\href{https://arxiv.org/abs/1910.02957}{{\ttfamily 1910.02957}}].

\bibitem{Dokuchaev:2019jqq}
V.~I. Dokuchaev and N.~O. Nazarova, \emph{{Silhouettes of invisible black
  holes}},  \href{https://arxiv.org/abs/1911.07695}{{\ttfamily 1911.07695}}.

\bibitem{Falcke:1999pj}
H.~Falcke, F.~Melia and E.~Agol, \emph{{Viewing the shadow of the black hole at
  the galactic center}}, \href{https://doi.org/10.1086/312423}{\emph{Astrophys.
  J.} {\bfseries 528} (2000) L13},
  [\href{https://arxiv.org/abs/astro-ph/9912263}{{\ttfamily
  astro-ph/9912263}}].

\bibitem{Doeleman:2009te}
S.~Doeleman et~al., \emph{{Imaging an Event Horizon: submm-VLBI of a Super
  Massive Black Hole}},  \href{https://arxiv.org/abs/0906.3899}{{\ttfamily
  0906.3899}}.

\bibitem{Akiyama:2019cqa}
{\scshape Event Horizon Telescope} collaboration, K.~Akiyama et~al.,
  \emph{{First M87 Event Horizon Telescope Results. I. The Shadow of the
  Supermassive Black Hole}},
  \href{https://doi.org/10.3847/2041-8213/ab0ec7}{\emph{Astrophys. J.}
  {\bfseries 875} (2019) L1},
  [\href{https://arxiv.org/abs/1906.11238}{{\ttfamily 1906.11238}}].

\bibitem{Akiyama:2019brx}
{\scshape Event Horizon Telescope} collaboration, K.~Akiyama et~al.,
  \emph{{First M87 Event Horizon Telescope Results. II. Array and
  Instrumentation}},
  \href{https://doi.org/10.3847/2041-8213/ab0c96}{\emph{Astrophys. J.}
  {\bfseries 875} (2019) L2},
  [\href{https://arxiv.org/abs/1906.11239}{{\ttfamily 1906.11239}}].

\bibitem{Akiyama:2019sww}
{\scshape Event Horizon Telescope} collaboration, K.~Akiyama et~al.,
  \emph{{First M87 Event Horizon Telescope Results. III. Data Processing and
  Calibration}},
  \href{https://doi.org/10.3847/2041-8213/ab0c57}{\emph{Astrophys. J.}
  {\bfseries 875} (2019) L3},
  [\href{https://arxiv.org/abs/1906.11240}{{\ttfamily 1906.11240}}].

\bibitem{Akiyama:2019bqs}
{\scshape Event Horizon Telescope} collaboration, K.~Akiyama et~al.,
  \emph{{First M87 Event Horizon Telescope Results. IV. Imaging the Central
  Supermassive Black Hole}},
  \href{https://doi.org/10.3847/2041-8213/ab0e85}{\emph{Astrophys. J.}
  {\bfseries 875} (2019) L4},
  [\href{https://arxiv.org/abs/1906.11241}{{\ttfamily 1906.11241}}].

\bibitem{Akiyama:2019fyp}
{\scshape Event Horizon Telescope} collaboration, K.~Akiyama et~al.,
  \emph{{First M87 Event Horizon Telescope Results. V. Physical Origin of the
  Asymmetric Ring}},
  \href{https://doi.org/10.3847/2041-8213/ab0f43}{\emph{Astrophys. J.}
  {\bfseries 875} (2019) L5},
  [\href{https://arxiv.org/abs/1906.11242}{{\ttfamily 1906.11242}}].

\bibitem{Akiyama:2019eap}
{\scshape Event Horizon Telescope} collaboration, K.~Akiyama et~al.,
  \emph{{First M87 Event Horizon Telescope Results. VI. The Shadow and Mass of
  the Central Black Hole}},
  \href{https://doi.org/10.3847/2041-8213/ab1141}{\emph{Astrophys. J.}
  {\bfseries 875} (2019) L6},
  [\href{https://arxiv.org/abs/1906.11243}{{\ttfamily 1906.11243}}].

\bibitem{Johannsen:2010ru}
T.~Johannsen and D.~Psaltis, \emph{{Testing the No-Hair Theorem with
  Observations in the Electromagnetic Spectrum: II. Black-Hole Images}},
  \href{https://doi.org/10.1088/0004-637X/718/1/446}{\emph{Astrophys. J.}
  {\bfseries 718} (2010) 446--454},
  [\href{https://arxiv.org/abs/1005.1931}{{\ttfamily 1005.1931}}].

\bibitem{Loeb:2013lfa}
A.~E. Broderick, T.~Johannsen, A.~Loeb and D.~Psaltis, \emph{{Testing the
  No-Hair Theorem with Event Horizon Telescope Observations of Sagittarius
  A*}}, \href{https://doi.org/10.1088/0004-637X/784/1/7}{\emph{Astrophys. J.}
  {\bfseries 784} (2014) 7}, [\href{https://arxiv.org/abs/1311.5564}{{\ttfamily
  1311.5564}}].

\bibitem{Johannsen:2015hib}
T.~Johannsen, A.~E. Broderick, P.~M. Plewa, S.~Chatzopoulos, S.~S. Doeleman,
  F.~Eisenhauer et~al., \emph{{Testing General Relativity with the Shadow Size
  of Sgr A*}},
  \href{https://doi.org/10.1103/PhysRevLett.116.031101}{\emph{Phys. Rev. Lett.}
  {\bfseries 116} (2016) 031101},
  [\href{https://arxiv.org/abs/1512.02640}{{\ttfamily 1512.02640}}].

\bibitem{Johannsen:2015mdd}
T.~Johannsen, \emph{{Sgr A* and General Relativity}},
  \href{https://doi.org/10.1088/0264-9381/33/11/113001}{\emph{Class. Quant.
  Grav.} {\bfseries 33} (2016) 113001},
  [\href{https://arxiv.org/abs/1512.03818}{{\ttfamily 1512.03818}}].

\bibitem{Psaltis:2018xkc}
D.~Psaltis, \emph{{Testing General Relativity with the Event Horizon
  Telescope}}, \href{https://doi.org/10.1007/s10714-019-2611-5}{\emph{Gen. Rel.
  Grav.} {\bfseries 51} (2019) 137},
  [\href{https://arxiv.org/abs/1806.09740}{{\ttfamily 1806.09740}}].

\bibitem{Israel:1967wq}
W.~Israel, \emph{{Event horizons in static vacuum space-times}},
  \href{https://doi.org/10.1103/PhysRev.164.1776}{\emph{Phys. Rev.} {\bfseries
  164} (1967) 1776--1779}.

\bibitem{Israel:1967za}
W.~Israel, \emph{{Event horizons in static electrovac space-times}},
  \href{https://doi.org/10.1007/BF01645859}{\emph{Commun. Math. Phys.}
  {\bfseries 8} (1968) 245--260}.

\bibitem{Carter:1971zc}
B.~Carter, \emph{{Axisymmetric Black Hole Has Only Two Degrees of Freedom}},
  \href{https://doi.org/10.1103/PhysRevLett.26.331}{\emph{Phys. Rev. Lett.}
  {\bfseries 26} (1971) 331--333}.

\bibitem{Moffat:2019uxp}
J.~W. Moffat and V.~T. Toth, \emph{{The masses and shadows of the black holes
  Sagittarius A* and M87 in modified gravity (MOG)}},
  \href{https://arxiv.org/abs/1904.04142}{{\ttfamily 1904.04142}}.

\bibitem{Nokhrina:2019sxv}
E.~E. Nokhrina, L.~I. Gurvits, V.~S. Beskin, M.~Nakamura, K.~Asada and K.~Hada,
  \emph{{M87 black hole mass and spin estimate through the position of the jet
  boundary shape break}},
  \href{https://doi.org/10.1093/mnras/stz2116}{\emph{Mon. Not. Roy. Astron.
  Soc.} {\bfseries 489} (2019) 1197--1205},
  [\href{https://arxiv.org/abs/1904.05665}{{\ttfamily 1904.05665}}].

\bibitem{Abdikamalov:2019ztb}
A.~B. Abdikamalov, A.~A. Abdujabbarov, D.~Ayzenberg, D.~Malafarina, C.~Bambi
  and B.~Ahmedov, \emph{{Black hole mimicker hiding in the shadow: Optical
  properties of the $\gamma$ metric}},
  \href{https://doi.org/10.1103/PhysRevD.100.024014}{\emph{Phys. Rev.}
  {\bfseries D100} (2019) 024014},
  [\href{https://arxiv.org/abs/1904.06207}{{\ttfamily 1904.06207}}].

\bibitem{Held:2019xde}
A.~Held, R.~Gold and A.~Eichhorn, \emph{{Asymptotic safety casts its shadow}},
  \href{https://doi.org/10.1088/1475-7516/2019/06/029}{\emph{JCAP} {\bfseries
  1906} (2019) 029}, [\href{https://arxiv.org/abs/1904.07133}{{\ttfamily
  1904.07133}}].

\bibitem{Wei:2019pjf}
S.-W. Wei, Y.-C. Zou, Y.-X. Liu and R.~B. Mann, \emph{{Curvature radius and
  Kerr black hole shadow}},
  \href{https://doi.org/10.1088/1475-7516/2019/08/030}{\emph{JCAP} {\bfseries
  1908} (2019) 030}, [\href{https://arxiv.org/abs/1904.07710}{{\ttfamily
  1904.07710}}].

\bibitem{Shaikh:2019fpu}
R.~Shaikh, \emph{{Black hole shadow in a general rotating spacetime obtained
  through Newman-Janis algorithm}},
  \href{https://doi.org/10.1103/PhysRevD.100.024028}{\emph{Phys. Rev.}
  {\bfseries D100} (2019) 024028},
  [\href{https://arxiv.org/abs/1904.08322}{{\ttfamily 1904.08322}}].

\bibitem{Tamburini:2019vrf}
F.~Tamburini, B.~Thidé and M.~Della~Valle, \emph{{Measurement of the spin of
  the M87 black hole from its observed twisted light}},
  \href{https://arxiv.org/abs/1904.07923}{{\ttfamily 1904.07923}}.

\bibitem{Davoudiasl:2019nlo}
H.~Davoudiasl and P.~B. Denton, \emph{{Ultralight Boson Dark Matter and Event
  Horizon Telescope Observations of M87*}},
  \href{https://doi.org/10.1103/PhysRevLett.123.021102}{\emph{Phys. Rev. Lett.}
  {\bfseries 123} (2019) 021102},
  [\href{https://arxiv.org/abs/1904.09242}{{\ttfamily 1904.09242}}].

\bibitem{Ovgun:2019yor}
A.~Övgün, I.~Sakalli and H.~Mutuk, \emph{{Quasinormal modes of Schwarzschild
  Black Hole and Damour-Solodukhin Wormhole via Feedforward Neural Network
  Method}},  \href{https://arxiv.org/abs/1904.09509}{{\ttfamily 1904.09509}}.

\bibitem{Bambi:2019tjh}
C.~Bambi, K.~Freese, S.~Vagnozzi and L.~Visinelli, \emph{{Testing the
  rotational nature of the supermassive object M87* from the circularity and
  size of its first image}},
  \href{https://doi.org/10.1103/PhysRevD.100.044057}{\emph{Phys. Rev.}
  {\bfseries D100} (2019) 044057},
  [\href{https://arxiv.org/abs/1904.12983}{{\ttfamily 1904.12983}}].

\bibitem{Nemmen:2019idv}
R.~Nemmen, \emph{{The Spin of M87*}},
  \href{https://doi.org/10.3847/2041-8213/ab2fd3}{\emph{Astrophys. J.}
  {\bfseries 880} (2019) L26},
  [\href{https://arxiv.org/abs/1905.02143}{{\ttfamily 1905.02143}}].

\bibitem{Churilova:2019jqx}
M.~S. Churilova, \emph{{Analytical quasinormal modes of spherically symmetric
  black holes in the eikonal regime}},
  \href{https://doi.org/10.1140/epjc/s10052-019-7146-0}{\emph{Eur. Phys. J.}
  {\bfseries C79} (2019) 629},
  [\href{https://arxiv.org/abs/1905.04536}{{\ttfamily 1905.04536}}].

\bibitem{Safarzadeh:2019imq}
M.~Safarzadeh, A.~Loeb and M.~Reid, \emph{{Constraining a black hole companion
  for M87* through imaging by the Event Horizon Telescope}},
  \href{https://doi.org/10.1093/mnrasl/slz108}{\emph{Mon. Not. Roy. Astron.
  Soc.} {\bfseries 488} (2019) L90--L93},
  [\href{https://arxiv.org/abs/1905.06835}{{\ttfamily 1905.06835}}].

\bibitem{Firouzjaee:2019aij}
J.~T. Firouzjaee and A.~Allahyari, \emph{{Black hole shadow with a cosmological
  constant for cosmological observers}},
  \href{https://doi.org/10.1140/epjc/s10052-019-7464-2}{\emph{Eur. Phys. J.}
  {\bfseries C79} (2019) 930},
  [\href{https://arxiv.org/abs/1905.07378}{{\ttfamily 1905.07378}}].

\bibitem{Konoplya:2019nzp}
R.~A. Konoplya, C.~Posada, Z.~Stuchlík and A.~Zhidenko, \emph{{Stable
  Schwarzschild stars as black-hole mimickers}},
  \href{https://doi.org/10.1103/PhysRevD.100.044027}{\emph{Phys. Rev.}
  {\bfseries D100} (2019) 044027},
  [\href{https://arxiv.org/abs/1905.08097}{{\ttfamily 1905.08097}}].

\bibitem{Kawashima:2019ljv}
T.~Kawashima, M.~Kino and K.~Akiyama, \emph{{Black Hole Spin Signature in the
  Black Hole Shadow of M87 in the Flaring State}},
  \href{https://doi.org/10.3847/1538-4357/ab19c0}{\emph{Astrophys. J.}
  {\bfseries 878} (2019) 27},
  [\href{https://arxiv.org/abs/1905.10717}{{\ttfamily 1905.10717}}].

\bibitem{Contreras:2019nih}
E.~Contreras, J.~M. Ramirez-Velasquez, A.~Rincón, G.~Panotopoulos and
  P.~Bargueño, \emph{{Black hole shadow of a rotating polytropic black hole by
  the Newman–Janis algorithm without complexification}},
  \href{https://doi.org/10.1140/epjc/s10052-019-7309-z}{\emph{Eur. Phys. J.}
  {\bfseries C79} (2019) 802},
  [\href{https://arxiv.org/abs/1905.11443}{{\ttfamily 1905.11443}}].

\bibitem{Bar:2019pnz}
N.~Bar, K.~Blum, T.~Lacroix and P.~Panci, \emph{{Looking for ultralight dark
  matter near supermassive black holes}},
  \href{https://doi.org/10.1088/1475-7516/2019/07/045}{\emph{JCAP} {\bfseries
  1907} (2019) 045}, [\href{https://arxiv.org/abs/1905.11745}{{\ttfamily
  1905.11745}}].

\bibitem{Jusufi:2019nrn}
K.~Jusufi, M.~Jamil, P.~Salucci, T.~Zhu and S.~Haroon, \emph{{Black Hole
  Surrounded by a Dark Matter Halo in the M87 Galactic Center and its
  Identification with Shadow Images}},
  \href{https://doi.org/10.1103/PhysRevD.100.044012}{\emph{Phys. Rev.}
  {\bfseries D100} (2019) 044012},
  [\href{https://arxiv.org/abs/1905.11803}{{\ttfamily 1905.11803}}].

\bibitem{Vagnozzi:2019apd}
S.~Vagnozzi and L.~Visinelli, \emph{{Hunting for extra dimensions in the shadow
  of M87*}}, \href{https://doi.org/10.1103/PhysRevD.100.024020}{\emph{Phys.
  Rev.} {\bfseries D100} (2019) 024020},
  [\href{https://arxiv.org/abs/1905.12421}{{\ttfamily 1905.12421}}].

\bibitem{Banerjee:2019cjk}
I.~Banerjee, B.~Mandal and S.~SenGupta, \emph{{Does black hole continuum
  spectrum signal higher curvature gravity in higher dimensions?}},
  \href{https://arxiv.org/abs/1905.12820}{{\ttfamily 1905.12820}}.

\bibitem{Roy:2019esk}
R.~Roy and U.~A. Yajnik, \emph{{Evolution of black hole shadow in the presence
  of ultralight bosons}},  \href{https://arxiv.org/abs/1906.03190}{{\ttfamily
  1906.03190}}.

\bibitem{Ali:2019khp}
M.~S. Ali and M.~Amir, \emph{{Shadow of rotating charged black hole with Weyl
  corrections}},  \href{https://arxiv.org/abs/1906.04146}{{\ttfamily
  1906.04146}}.

\bibitem{Long:2019nox}
F.~Long, J.~Wang, S.~Chen and J.~Jing, \emph{{Shadow of a rotating squashed
  Kaluza-Klein black hole}},
  \href{https://doi.org/10.1007/JHEP10(2019)269}{\emph{JHEP} {\bfseries 10}
  (2019) 269}, [\href{https://arxiv.org/abs/1906.04456}{{\ttfamily
  1906.04456}}].

\bibitem{Zhu:2019ura}
T.~Zhu, Q.~Wu, M.~Jamil and K.~Jusufi, \emph{{Shadows and deflection angle of
  charged and slowly rotating black holes in Einstein-Æther theory}},
  \href{https://doi.org/10.1103/PhysRevD.100.044055}{\emph{Phys. Rev.}
  {\bfseries D100} (2019) 044055},
  [\href{https://arxiv.org/abs/1906.05673}{{\ttfamily 1906.05673}}].

\bibitem{Contreras:2019cmf}
E.~Contreras, A.~Rincón, G.~Panotopoulos, P.~Bargueño and B.~Koch,
  \emph{{Black hole shadow of a rotating scale--dependent black hole}},
  \href{https://arxiv.org/abs/1906.06990}{{\ttfamily 1906.06990}}.

\bibitem{Dokuchaev:2019pcx}
V.~I. Dokuchaev and N.~O. Nazarova, \emph{{The brightest point in accretion
  disk and black hole spin: implication to the image of black hole M87*}},
  \href{https://doi.org/10.3390/universe5080183}{\emph{Universe} {\bfseries 5}
  (2019) 183}, [\href{https://arxiv.org/abs/1906.07171}{{\ttfamily
  1906.07171}}].

\bibitem{Qi:2019zdk}
J.-Z. Qi and X.~Zhang, \emph{{A new cosmological probe from supermassive black
  hole shadows}},  \href{https://arxiv.org/abs/1906.10825}{{\ttfamily
  1906.10825}}.

\bibitem{Wang:2019tto}
L.-F. Wang, Z.-W. Zhao, J.-F. Zhang and X.~Zhang, \emph{{A preliminary forecast
  for cosmological parameter estimation with gravitational-wave standard sirens
  from TianQin}},  \href{https://arxiv.org/abs/1907.01838}{{\ttfamily
  1907.01838}}.

\bibitem{Konoplya:2019goy}
R.~A. Konoplya and A.~Zhidenko, \emph{{Analytical representation for metrics of
  scalarized Einstein-Maxwell black holes and their shadows}},
  \href{https://doi.org/10.1103/PhysRevD.100.044015}{\emph{Phys. Rev.}
  {\bfseries D100} (2019) 044015},
  [\href{https://arxiv.org/abs/1907.05551}{{\ttfamily 1907.05551}}].

\bibitem{Roy:2019hqf}
P.~Roy and R.~Biswas, \emph{{Accretion onto Quintessence Contaminated Black
  Hole}},  \href{https://arxiv.org/abs/1907.06719}{{\ttfamily 1907.06719}}.

\bibitem{Pavlovic:2019rim}
P.~Pavlović, A.~Saveliev and M.~Sossich, \emph{{Influence of the Vacuum
  Polarization Effect on the Motion of Charged Particles in the Magnetic Field
  around a Schwarzschild Black Hole}},
  \href{https://doi.org/10.1103/PhysRevD.100.084033}{\emph{Phys. Rev.}
  {\bfseries D100} (2019) 084033},
  [\href{https://arxiv.org/abs/1908.01888}{{\ttfamily 1908.01888}}].

\bibitem{Biswas:2019gia}
R.~Biswas and S.~Dutta, \emph{{Threshold Drop in Accretion Density if Dark
  Energy is Accreting onto a Supermassive Black Hole}},
  \href{https://doi.org/10.1140/epjc/s10052-019-7258-6}{\emph{Eur. Phys. J.}
  {\bfseries C79} (2019) 742},
  [\href{https://arxiv.org/abs/1908.04268}{{\ttfamily 1908.04268}}].

\bibitem{Wang:2019skw}
M.~Wang, S.~Chen and J.~Jing, \emph{{Effect of gravitational wave on shadow of
  a Schwarzschild black hole}},
  \href{https://arxiv.org/abs/1908.04527}{{\ttfamily 1908.04527}}.

\bibitem{Nalewajko:2019mxh}
K.~Nalewajko, M.~Sikora and A.~Różańska, \emph{{On the orientation of the
  crescent image of M87*}},  \href{https://arxiv.org/abs/1908.10376}{{\ttfamily
  1908.10376}}.

\bibitem{Tian:2019yhn}
S.~X. Tian and Z.-H. Zhu, \emph{{Testing the Schwarzschild metric in a strong
  field region with the Event Horizon Telescope}},
  \href{https://doi.org/10.1103/PhysRevD.100.064011}{\emph{Phys. Rev.}
  {\bfseries D100} (2019) 064011},
  [\href{https://arxiv.org/abs/1908.11794}{{\ttfamily 1908.11794}}].

\bibitem{Cunha:2019ikd}
P.~V.~P. Cunha, C.~A.~R. Herdeiro and E.~Radu, \emph{{EHT constraint on the
  ultralight scalar hair of the M87 supermassive black hole}},
  \href{https://arxiv.org/abs/1909.08039}{{\ttfamily 1909.08039}}.

\bibitem{Banerjee:2019nnj}
I.~Banerjee, S.~Chakraborty and S.~SenGupta, \emph{{Silhouette of M87*: A New
  Window to Peek into the World of Hidden Dimensions}},
  \href{https://arxiv.org/abs/1909.09385}{{\ttfamily 1909.09385}}.

\bibitem{Shaikh:2019hbm}
R.~Shaikh and P.~S. Joshi, \emph{{Can we distinguish black holes from naked
  singularities by the images of their accretion disks?}},
  \href{https://doi.org/10.1088/1475-7516/2019/10/064}{\emph{JCAP} {\bfseries
  1910} (2019) 064}, [\href{https://arxiv.org/abs/1909.10322}{{\ttfamily
  1909.10322}}].

\bibitem{Vrba:2019vqh}
J.~Vrba, A.~Abdujabbarov, A.~Tursunov, B.~Ahmedov and Z.~Stuchlík,
  \emph{{Particle motion around generic black holes coupled to non-linear
  electrodynamics}},
  \href{https://doi.org/10.1140/epjc/s10052-019-7286-2}{\emph{Eur. Phys. J.}
  {\bfseries C79} (2019) 778},
  [\href{https://arxiv.org/abs/1909.12026}{{\ttfamily 1909.12026}}].

\bibitem{Kumar:2019ghw}
R.~Kumar, S.~G. Ghosh and A.~Wang, \emph{{Shadow cast and deflection of light
  by charged rotating regular black holes}},
  \href{https://arxiv.org/abs/1912.05154}{{\ttfamily 1912.05154}}.

\bibitem{Hawking:1969sw}
S.~W. Hawking and R.~Penrose, \emph{{The Singularities of gravitational
  collapse and cosmology}},
  \href{https://doi.org/10.1098/rspa.1970.0021}{\emph{Proc. Roy. Soc. Lond.}
  {\bfseries A314} (1970) 529--548}.

\bibitem{Senovilla:2018aav}
J.~M.~M. Senovilla, \emph{{Singularity Theorems and Their Consequences}},
  \href{https://arxiv.org/abs/1801.04912}{{\ttfamily 1801.04912}}.

\bibitem{Penrose:1969pc}
R.~Penrose, \emph{{Gravitational collapse: The role of general relativity}},
  \href{https://doi.org/10.1023/A:1016578408204}{\emph{Riv. Nuovo Cim.}
  {\bfseries 1} (1969) 252--276}.

\bibitem{Wald:1997wa}
R.~M. Wald, \emph{{Gravitational collapse and cosmic censorship}},  in
  \emph{{Black Holes, Gravitational Radiation and the Universe: Essays in Honor
  of C.V. Vishveshwara}}, pp.~69--85, 1997,
  \href{https://arxiv.org/abs/gr-qc/9710068}{{\ttfamily gr-qc/9710068}},
  \href{https://doi.org/10.1007/978-94-017-0934-7_5}{DOI}.

\bibitem{Bardeen:1968ghw}
J.~Bardeen, \emph{{Non-singular general-relativistic gravitational collapse}},
  in \emph{{Proceedings, 5th International Conference on Gravitation and the
  theory of relativity: Tbilisi, USSR, December 9-13, 1968}}, p.~174, 1968.

\bibitem{Borde:1996df}
A.~Borde, \emph{{Regular black holes and topology change}},
  \href{https://doi.org/10.1103/PhysRevD.55.7615}{\emph{Phys. Rev.} {\bfseries
  D55} (1997) 7615--7617},
  [\href{https://arxiv.org/abs/gr-qc/9612057}{{\ttfamily gr-qc/9612057}}].

\bibitem{AyonBeato:1998ub}
E.~Ayon-Beato and A.~Garcia, \emph{{Regular black hole in general relativity
  coupled to nonlinear electrodynamics}},
  \href{https://doi.org/10.1103/PhysRevLett.80.5056}{\emph{Phys. Rev. Lett.}
  {\bfseries 80} (1998) 5056--5059},
  [\href{https://arxiv.org/abs/gr-qc/9911046}{{\ttfamily gr-qc/9911046}}].

\bibitem{AyonBeato:1999rg}
E.~Ayon-Beato and A.~Garcia, \emph{{New regular black hole solution from
  nonlinear electrodynamics}},
  \href{https://doi.org/10.1016/S0370-2693(99)01038-2}{\emph{Phys. Lett.}
  {\bfseries B464} (1999) 25},
  [\href{https://arxiv.org/abs/hep-th/9911174}{{\ttfamily hep-th/9911174}}].

\bibitem{Bronnikov:2000yz}
K.~A. Bronnikov, \emph{{Comment on `Regular black hole in general relativity
  coupled to nonlinear electrodynamics'}},
  \href{https://doi.org/10.1103/PhysRevLett.85.4641}{\emph{Phys. Rev. Lett.}
  {\bfseries 85} (2000) 4641}.

\bibitem{bur1}
C.~P. Burgess, R.~Easther, A.~Mazumdar, D.~F. Mota and T.~Multamaki,
  \emph{{Multiple inflation, cosmic string networks and the string landscape}},
  \href{https://doi.org/10.1088/1126-6708/2005/05/067}{\emph{JHEP} {\bfseries
  05} (2005) 067}, [\href{https://arxiv.org/abs/hep-th/0501125}{{\ttfamily
  hep-th/0501125}}].

\bibitem{Hayward:2005gi}
S.~A. Hayward, \emph{{Formation and evaporation of regular black holes}},
  \href{https://doi.org/10.1103/PhysRevLett.96.031103}{\emph{Phys. Rev. Lett.}
  {\bfseries 96} (2006) 031103},
  [\href{https://arxiv.org/abs/gr-qc/0506126}{{\ttfamily gr-qc/0506126}}].

\bibitem{Bronnikov:2005gm}
K.~A. Bronnikov and J.~C. Fabris, \emph{{Regular phantom black holes}},
  \href{https://doi.org/10.1103/PhysRevLett.96.251101}{\emph{Phys. Rev. Lett.}
  {\bfseries 96} (2006) 251101},
  [\href{https://arxiv.org/abs/gr-qc/0511109}{{\ttfamily gr-qc/0511109}}].

\bibitem{Iso:2006ut}
S.~Iso, H.~Umetsu and F.~Wilczek, \emph{{Anomalies, Hawking radiations and
  regularity in rotating black holes}},
  \href{https://doi.org/10.1103/PhysRevD.74.044017}{\emph{Phys. Rev.}
  {\bfseries D74} (2006) 044017},
  [\href{https://arxiv.org/abs/hep-th/0606018}{{\ttfamily hep-th/0606018}}].

\bibitem{Berej:2006cc}
W.~Berej, J.~Matyjasek, D.~Tryniecki and M.~Woronowicz, \emph{{Regular black
  holes in quadratic gravity}},
  \href{https://doi.org/10.1007/s10714-006-0270-9}{\emph{Gen. Rel. Grav.}
  {\bfseries 38} (2006) 885--906},
  [\href{https://arxiv.org/abs/hep-th/0606185}{{\ttfamily hep-th/0606185}}].

\bibitem{bur3}
B.~Li, D.~F. Mota and D.~J. Shaw, \emph{{Microscopic and Macroscopic Behaviors
  of Palatini Modified Gravity Theories}},
  \href{https://doi.org/10.1103/PhysRevD.78.064018}{\emph{Phys. Rev.}
  {\bfseries D78} (2008) 064018},
  [\href{https://arxiv.org/abs/0805.3428}{{\ttfamily 0805.3428}}].

\bibitem{bur2}
A.~De~Felice, D.~F. Mota and S.~Tsujikawa, \emph{{Matter instabilities in
  general Gauss-Bonnet gravity}},
  \href{https://doi.org/10.1103/PhysRevD.81.023532}{\emph{Phys. Rev.}
  {\bfseries D81} (2010) 023532},
  [\href{https://arxiv.org/abs/0911.1811}{{\ttfamily 0911.1811}}].

\bibitem{Bronnikov:2012ch}
K.~A. Bronnikov, R.~A. Konoplya and A.~Zhidenko, \emph{{Instabilities of
  wormholes and regular black holes supported by a phantom scalar field}},
  \href{https://doi.org/10.1103/PhysRevD.86.024028}{\emph{Phys. Rev.}
  {\bfseries D86} (2012) 024028},
  [\href{https://arxiv.org/abs/1205.2224}{{\ttfamily 1205.2224}}].

\bibitem{Rinaldi:2012vy}
M.~Rinaldi, \emph{{Black holes with non-minimal derivative coupling}},
  \href{https://doi.org/10.1103/PhysRevD.86.084048}{\emph{Phys. Rev.}
  {\bfseries D86} (2012) 084048},
  [\href{https://arxiv.org/abs/1208.0103}{{\ttfamily 1208.0103}}].

\bibitem{Bambi:2013ufa}
C.~Bambi and L.~Modesto, \emph{{Rotating regular black holes}},
  \href{https://doi.org/10.1016/j.physletb.2013.03.025}{\emph{Phys. Lett.}
  {\bfseries B721} (2013) 329--334},
  [\href{https://arxiv.org/abs/1302.6075}{{\ttfamily 1302.6075}}].

\bibitem{Sebastiani:2013fsa}
L.~Sebastiani, D.~Momeni, R.~Myrzakulov and S.~D. Odintsov,
  \emph{{Instabilities and (anti)-evaporation of Schwarzschild–de Sitter
  black holes in modified gravity}},
  \href{https://doi.org/10.1103/PhysRevD.88.104022}{\emph{Phys. Rev.}
  {\bfseries D88} (2013) 104022},
  [\href{https://arxiv.org/abs/1305.4231}{{\ttfamily 1305.4231}}].

\bibitem{Toshmatov:2014nya}
B.~Toshmatov, B.~Ahmedov, A.~Abdujabbarov and Z.~Stuchlik, \emph{{Rotating
  Regular Black Hole Solution}},
  \href{https://doi.org/10.1103/PhysRevD.89.104017}{\emph{Phys. Rev.}
  {\bfseries D89} (2014) 104017},
  [\href{https://arxiv.org/abs/1404.6443}{{\ttfamily 1404.6443}}].

\bibitem{Johannsen:2015pca}
T.~Johannsen, \emph{{Regular Black Hole Metric with Three Constants of
  Motion}}, \href{https://doi.org/10.1103/PhysRevD.88.044002}{\emph{Phys. Rev.}
  {\bfseries D88} (2013) 044002},
  [\href{https://arxiv.org/abs/1501.02809}{{\ttfamily 1501.02809}}].

\bibitem{Myrzakulov:2015sea}
R.~Myrzakulov and L.~Sebastiani, \emph{{Spherically symmetric static vacuum
  solutions in Mimetic gravity}},
  \href{https://doi.org/10.1007/s10714-015-1930-4}{\emph{Gen. Rel. Grav.}
  {\bfseries 47} (2015) 89},
  [\href{https://arxiv.org/abs/1503.04293}{{\ttfamily 1503.04293}}].

\bibitem{Myrzakulov:2015qaa}
R.~Myrzakulov, L.~Sebastiani and S.~Vagnozzi, \emph{{Inflation in $f(R,\phi )$
  -theories and mimetic gravity scenario}},
  \href{https://doi.org/10.1140/epjc/s10052-015-3672-6}{\emph{Eur. Phys. J.}
  {\bfseries C75} (2015) 444},
  [\href{https://arxiv.org/abs/1504.07984}{{\ttfamily 1504.07984}}].

\bibitem{Myrzakulov:2015kda}
R.~Myrzakulov, L.~Sebastiani, S.~Vagnozzi and S.~Zerbini, \emph{{Static
  spherically symmetric solutions in mimetic gravity: rotation curves and
  wormholes}},
  \href{https://doi.org/10.1088/0264-9381/33/12/125005}{\emph{Class. Quant.
  Grav.} {\bfseries 33} (2016) 125005},
  [\href{https://arxiv.org/abs/1510.02284}{{\ttfamily 1510.02284}}].

\bibitem{Abdujabbarov:2016hnw}
A.~Abdujabbarov, M.~Amir, B.~Ahmedov and S.~G. Ghosh, \emph{{Shadow of rotating
  regular black holes}},
  \href{https://doi.org/10.1103/PhysRevD.93.104004}{\emph{Phys. Rev.}
  {\bfseries D93} (2016) 104004},
  [\href{https://arxiv.org/abs/1604.03809}{{\ttfamily 1604.03809}}].

\bibitem{Fan:2016hvf}
Z.-Y. Fan and X.~Wang, \emph{{Construction of Regular Black Holes in General
  Relativity}}, \href{https://doi.org/10.1103/PhysRevD.94.124027}{\emph{Phys.
  Rev.} {\bfseries D94} (2016) 124027},
  [\href{https://arxiv.org/abs/1610.02636}{{\ttfamily 1610.02636}}].

\bibitem{Sebastiani:2016ras}
L.~Sebastiani, S.~Vagnozzi and R.~Myrzakulov, \emph{{Mimetic gravity: a review
  of recent developments and applications to cosmology and astrophysics}},
  \href{https://doi.org/10.1155/2017/3156915}{\emph{Adv. High Energy Phys.}
  {\bfseries 2017} (2017) 3156915},
  [\href{https://arxiv.org/abs/1612.08661}{{\ttfamily 1612.08661}}].

\bibitem{Toshmatov:2017zpr}
B.~Toshmatov, Z.~Stuchlík and B.~Ahmedov, \emph{{Generic rotating regular
  black holes in general relativity coupled to nonlinear electrodynamics}},
  \href{https://doi.org/10.1103/PhysRevD.95.084037}{\emph{Phys. Rev.}
  {\bfseries D95} (2017) 084037},
  [\href{https://arxiv.org/abs/1704.07300}{{\ttfamily 1704.07300}}].

\bibitem{Chinaglia:2017uqd}
S.~Chinaglia and S.~Zerbini, \emph{{A note on singular and non-singular black
  holes}}, \href{https://doi.org/10.1007/s10714-017-2235-6}{\emph{Gen. Rel.
  Grav.} {\bfseries 49} (2017) 75},
  [\href{https://arxiv.org/abs/1704.08516}{{\ttfamily 1704.08516}}].

\bibitem{Chinaglia:2017wih}
S.~Chinaglia, \emph{{A model of regular black hole satisfying the Weak Energy
  Condition}},  \href{https://arxiv.org/abs/1707.02795}{{\ttfamily
  1707.02795}}.

\bibitem{Colleaux:2017ibe}
A.~Colléaux, S.~Chinaglia and S.~Zerbini, \emph{{Nonpolynomial Lagrangian
  approach to regular black holes}},
  \href{https://doi.org/10.1142/S0218271818300021}{\emph{Int. J. Mod. Phys.}
  {\bfseries D27} (2018) 1830002},
  [\href{https://arxiv.org/abs/1712.03730}{{\ttfamily 1712.03730}}].

\bibitem{Jusufi:2018jof}
K.~Jusufi, A.~Övgün, J.~Saavedra, Y.~Vásquez and P.~A. González,
  \emph{{Deflection of light by rotating regular black holes using the
  Gauss-Bonnet theorem}},
  \href{https://doi.org/10.1103/PhysRevD.97.124024}{\emph{Phys. Rev.}
  {\bfseries D97} (2018) 124024},
  [\href{https://arxiv.org/abs/1804.00643}{{\ttfamily 1804.00643}}].

\bibitem{Chinaglia:2018gvf}
S.~Chinaglia, \emph{{A no-go theorem for regular black holes}},
  \href{https://arxiv.org/abs/1805.03899}{{\ttfamily 1805.03899}}.

\bibitem{Ovgun:2019wej}
A.~Övgün, \emph{{Weak field deflection angle by regular black holes with
  cosmic strings using the Gauss-Bonnet theorem}},
  \href{https://doi.org/10.1103/PhysRevD.99.104075}{\emph{Phys. Rev.}
  {\bfseries D99} (2019) 104075},
  [\href{https://arxiv.org/abs/1902.04411}{{\ttfamily 1902.04411}}].

\bibitem{Han:2019lfs}
Y.-W. Han, M.-J. Lan and X.-X. Zeng, \emph{{Thermodynamics and weak cosmic
  censorship conjecture in (2+1)-dimensional regular black hole with nonlinear
  electrodynamics sources}},
  \href{https://arxiv.org/abs/1903.03764}{{\ttfamily 1903.03764}}.

\bibitem{Rodrigues:2019xrc}
M.~E. Rodrigues and M.~V. de~S.~Silva, \emph{{Regular multihorizon black holes
  in $f(G)$ gravity with nonlinear electrodynamics}},
  \href{https://doi.org/10.1103/PhysRevD.99.124010}{\emph{Phys. Rev.}
  {\bfseries D99} (2019) 124010},
  [\href{https://arxiv.org/abs/1906.06168}{{\ttfamily 1906.06168}}].

\bibitem{Panotopoulos:2019qjk}
G.~Panotopoulos and A.~Rincón, \emph{{Quasinormal modes of regular black holes
  with non linear-Electrodynamical sources}},
  \href{https://doi.org/10.1140/epjp/i2019-12686-x}{\emph{Eur. Phys. J. Plus}
  {\bfseries 134} (2019) 300},
  [\href{https://arxiv.org/abs/1904.10847}{{\ttfamily 1904.10847}}].

\bibitem{Jusufi:2019caq}
K.~Jusufi, M.~Jamil, H.~Chakrabarty, C.~Bambi and A.~Wang, \emph{{Rotating
  regular black holes in conformal massive gravity}},
  \href{https://arxiv.org/abs/1911.07520}{{\ttfamily 1911.07520}}.

\bibitem{Gorji:2019hog}
M.~A. Gorji, A.~Allahyari, M.~Khodadi and H.~Firouzjahi, \emph{{Mimetic Black
  Holes}},  \href{https://arxiv.org/abs/1912.04636}{{\ttfamily 1912.04636}}.

\bibitem{Stuchlik:2014qja}
Z.~Stuchlík and J.~Schee, \emph{{Circular geodesic of Bardeen and
  Ayon–Beato–Garcia regular black-hole and no-horizon spacetimes}},
  \href{https://doi.org/10.1142/S0218271815500200}{\emph{Int. J. Mod. Phys.}
  {\bfseries D24} (2014) 1550020},
  [\href{https://arxiv.org/abs/1501.00015}{{\ttfamily 1501.00015}}].

\bibitem{Schee:2015nua}
J.~Schee and Z.~Stuchlik, \emph{{Gravitational lensing and ghost images in the
  regular Bardeen no-horizon spacetimes}},
  \href{https://doi.org/10.1088/1475-7516/2015/06/048}{\emph{JCAP} {\bfseries
  1506} (2015) 048}, [\href{https://arxiv.org/abs/1501.00835}{{\ttfamily
  1501.00835}}].

\bibitem{Schee:2016mjd}
J.~Schee and Z.~Stuchlík, \emph{{Profiled spectral lines generated by
  Keplerian discs orbiting in the Bardeen and Ayòn-Beato–Garcìa
  spacetimes}},
  \href{https://doi.org/10.1088/0264-9381/33/8/085004}{\emph{Class. Quant.
  Grav.} {\bfseries 33} (2016) 085004},
  [\href{https://arxiv.org/abs/1604.00632}{{\ttfamily 1604.00632}}].

\bibitem{Stuchlik:2019uvf}
Z.~Stuchlík and J.~Schee, \emph{{Shadow of the regular Bardeen black holes and
  comparison of the motion of photons and neutrinos}},
  \href{https://doi.org/10.1140/epjc/s10052-019-6543-8}{\emph{Eur. Phys. J.}
  {\bfseries C79} (2019) 44}.

\bibitem{Schee:2019gki}
J.~Schee and Z.~Stuchlik, \emph{{Profiled spectral lines of Keplerian rings
  orbiting in the regular Bardeen black hole spacetimes}},
  \href{https://doi.org/10.1140/epjc/s10052-019-7420-1}{\emph{Eur. Phys. J.}
  {\bfseries C79} (2019) 988},
  [\href{https://arxiv.org/abs/1908.07197}{{\ttfamily 1908.07197}}].

\bibitem{Born:1934gh}
M.~Born and L.~Infeld, \emph{{Foundations of the new field theory}},
  \href{https://doi.org/10.1098/rspa.1934.0059}{\emph{Proc. Roy. Soc. Lond.}
  {\bfseries A144} (1934) 425--451}.

\bibitem{Heisenberg:1935qt}
W.~Heisenberg and H.~Euler, \emph{{Consequences of Dirac's theory of
  positrons}}, \href{https://doi.org/10.1007/BF01343663,
  10.1007/978-3-642-70078-1_9}{\emph{Z. Phys.} {\bfseries 98} (1936) 714--732},
  [\href{https://arxiv.org/abs/physics/0605038}{{\ttfamily physics/0605038}}].

\bibitem{Bronnikov:2000vy}
K.~A. Bronnikov, \emph{{Regular magnetic black holes and monopoles from
  nonlinear electrodynamics}},
  \href{https://doi.org/10.1103/PhysRevD.63.044005}{\emph{Phys. Rev.}
  {\bfseries D63} (2001) 044005},
  [\href{https://arxiv.org/abs/gr-qc/0006014}{{\ttfamily gr-qc/0006014}}].

\bibitem{Stehle:1966wii}
P.~Stehle and P.~G. DeBaryshe, \emph{{Quantum Electrodynamics and the
  Correspondence Principle}},
  \href{https://doi.org/10.1103/PhysRev.152.1135}{\emph{Phys. Rev.} {\bfseries
  152} (1966) 1135}.

\bibitem{Fradkin:1985qd}
E.~S. Fradkin and A.~A. Tseytlin, \emph{{Nonlinear Electrodynamics from
  Quantized Strings}},
  \href{https://doi.org/10.1016/0370-2693(85)90205-9}{\emph{Phys. Lett.}
  {\bfseries 163B} (1985) 123--130}.

\bibitem{Tseytlin:1986ti}
A.~A. Tseytlin, \emph{{Vector Field Effective Action in the Open Superstring
  Theory}}, \href{https://doi.org/10.1016/0550-3213(86)90303-2,
  10.1016/0550-3213(87)90500-1}{\emph{Nucl. Phys.} {\bfseries B276} (1986)
  391}.

\bibitem{Bern:1993tz}
Z.~Bern and A.~G. Morgan, \emph{{Supersymmetry relations between contributions
  to one loop gauge boson amplitudes}},
  \href{https://doi.org/10.1103/PhysRevD.49.6155}{\emph{Phys. Rev.} {\bfseries
  D49} (1994) 6155--6163},
  [\href{https://arxiv.org/abs/hep-ph/9312218}{{\ttfamily hep-ph/9312218}}].

\bibitem{Dunne:2004nc}
G.~V. Dunne, \emph{{Heisenberg-Euler effective Lagrangians: Basics and
  extensions}},  in \emph{From fields to strings: Circumnavigating theoretical
  physics. Ian Kogan memorial collection (3 volume set)} (M.~Shifman,
  A.~Vainshtein and J.~Wheater, eds.), pp.~445--522.
\newblock 2004.
\newblock \href{https://arxiv.org/abs/hep-th/0406216}{{\ttfamily
  hep-th/0406216}}.
\newblock \href{https://doi.org/10.1142/9789812775344_0014}{DOI}.

\bibitem{Jacobson:2018kso}
T.~N. Jacobson and T.~Ter~Veldhuis, \emph{{Nonperturbative one-loop effective
  action for QED with Yukawa couplings}},
  \href{https://doi.org/10.1142/S0217751X18501579}{\emph{Int. J. Mod. Phys.}
  {\bfseries A33} (2018) 1850157},
  [\href{https://arxiv.org/abs/1806.04040}{{\ttfamily 1806.04040}}].

\bibitem{Dey:2004yt}
T.~K. Dey, \emph{{Born-Infeld black holes in the presence of a cosmological
  constant}}, \href{https://doi.org/10.1016/j.physletb.2004.06.047}{\emph{Phys.
  Lett.} {\bfseries B595} (2004) 484--490},
  [\href{https://arxiv.org/abs/hep-th/0406169}{{\ttfamily hep-th/0406169}}].

\bibitem{Cai:2004eh}
R.-G. Cai, D.-W. Pang and A.~Wang, \emph{{Born-Infeld black holes in (A)dS
  spaces}}, \href{https://doi.org/10.1103/PhysRevD.70.124034}{\emph{Phys. Rev.}
  {\bfseries D70} (2004) 124034},
  [\href{https://arxiv.org/abs/hep-th/0410158}{{\ttfamily hep-th/0410158}}].

\bibitem{Stefanov:2007qw}
I.~Z. Stefanov, S.~S. Yazadjiev and M.~D. Todorov, \emph{{Scalar-tensor black
  holes coupled to Born-Infeld nonlinear electrodynamics}},
  \href{https://doi.org/10.1103/PhysRevD.75.084036}{\emph{Phys. Rev.}
  {\bfseries D75} (2007) 084036},
  [\href{https://arxiv.org/abs/0704.3784}{{\ttfamily 0704.3784}}].

\bibitem{HabibMazharimousavi:2008dm}
S.~Habib~Mazharimousavi, M.~Halilsoy and Z.~Amirabi, \emph{{New Non-Abelian
  Black Hole Solutions in Born-Infeld gravity}},
  \href{https://doi.org/10.1103/PhysRevD.78.064050}{\emph{Phys. Rev.}
  {\bfseries D78} (2008) 064050},
  [\href{https://arxiv.org/abs/0806.4614}{{\ttfamily 0806.4614}}].

\bibitem{Stefanov:2009qd}
I.~Z. Stefanov, S.~S. Yazadjiev, D.~A. Georgieva and M.~D. Todorov,
  \emph{{Born-Infeld black holes coupled to a massive scalar field}},
  \href{https://doi.org/10.1142/S0218271811020469}{\emph{Int. J. Mod. Phys.}
  {\bfseries D20} (2011) 2471--2496},
  [\href{https://arxiv.org/abs/0909.0196}{{\ttfamily 0909.0196}}].

\bibitem{Ghodsi:2010ev}
A.~Ghodsi and D.~M. Yekta, \emph{{Black Holes in Born-Infeld Extended New
  Massive Gravity}},
  \href{https://doi.org/10.1103/PhysRevD.83.104004}{\emph{Phys. Rev.}
  {\bfseries D83} (2011) 104004},
  [\href{https://arxiv.org/abs/1010.2434}{{\ttfamily 1010.2434}}].

\bibitem{Allahverdizadeh:2013oha}
M.~Allahverdizadeh, J.~P.~S. Lemos and A.~Sheykhi, \emph{{Extremal Myers-Perry
  black holes coupled to Born-Infeld electrodynamics in five dimensions}},
  \href{https://doi.org/10.1103/PhysRevD.87.084002}{\emph{Phys. Rev.}
  {\bfseries D87} (2013) 084002},
  [\href{https://arxiv.org/abs/1302.5079}{{\ttfamily 1302.5079}}].

\bibitem{Allaverdizadeh:2013rua}
M.~Allaverdizadeh, S.~H. Hendi, J.~P.~S. Lemos and A.~Sheykhi, \emph{{Extremal
  Myers-Perry black holes coupled to Born-Infeld electrodynamics in odd
  dimensions}}, \href{https://doi.org/10.1142/S0218271814500321}{\emph{Int. J.
  Mod. Phys.} {\bfseries D23} (2014) 1450032},
  [\href{https://arxiv.org/abs/1304.0836}{{\ttfamily 1304.0836}}].

\bibitem{Olmo:2013gqa}
G.~J. Olmo, D.~Rubiera-Garcia and H.~Sanchis-Alepuz, \emph{{Geonic black holes
  and remnants in Eddington-inspired Born-Infeld gravity}},
  \href{https://doi.org/10.1140/epjc/s10052-014-2804-8}{\emph{Eur. Phys. J.}
  {\bfseries C74} (2014) 2804},
  [\href{https://arxiv.org/abs/1311.0815}{{\ttfamily 1311.0815}}].

\bibitem{Wu:2016hry}
J.-P. Wu, \emph{{Holographic fermionic spectrum from Born–Infeld AdS black
  hole}}, \href{https://doi.org/10.1016/j.physletb.2016.05.049}{\emph{Phys.
  Lett.} {\bfseries B758} (2016) 440--448},
  [\href{https://arxiv.org/abs/1705.06707}{{\ttfamily 1705.06707}}].

\bibitem{Bambi:2016xme}
C.~Bambi, D.~Rubiera-Garcia and Y.~Wang, \emph{{Black hole solutions in
  functional extensions of Born-Infeld gravity}},
  \href{https://doi.org/10.1103/PhysRevD.94.064002}{\emph{Phys. Rev.}
  {\bfseries D94} (2016) 064002},
  [\href{https://arxiv.org/abs/1608.04873}{{\ttfamily 1608.04873}}].

\bibitem{Chen:2017ify}
C.-Y. Chen, M.~Bouhmadi-López and P.~Chen, \emph{{Black hole solutions in
  mimetic Born-Infeld gravity}},
  \href{https://doi.org/10.1140/epjc/s10052-018-5556-z}{\emph{Eur. Phys. J.}
  {\bfseries C78} (2018) 59},
  [\href{https://arxiv.org/abs/1710.10638}{{\ttfamily 1710.10638}}].

\bibitem{Boehmer:2019uxv}
C.~G. Böhmer and F.~Fiorini, \emph{{The regular black hole in four dimensional
  Born–Infeld gravity}},
  \href{https://doi.org/10.1088/1361-6382/ab1e8d}{\emph{Class. Quant. Grav.}
  {\bfseries 36} (2019) 12LT01},
  [\href{https://arxiv.org/abs/1901.02965}{{\ttfamily 1901.02965}}].

\bibitem{Kumar:2019zbp}
N.~Kumar, S.~Bhattacharyya and S.~Gangopadhyay, \emph{{Phase transitions in
  Born-Infeld AdS black holes in D-dimensions}},
  \href{https://arxiv.org/abs/1904.13059}{{\ttfamily 1904.13059}}.

\bibitem{Li:2019qbw}
H.~Li and J.~Wang, \emph{{Black hole magnetospheres in the Born-Infeld
  theory}},  \href{https://arxiv.org/abs/1908.11104}{{\ttfamily 1908.11104}}.

\bibitem{Bronnikov:1979ex}
K.~A. Bronnikov, V.~N. Melnikov, G.~N. Shikin and K.~P. Staniukowicz,
  \emph{{SCALAR, ELECTROMAGNETIC, AND GRAVITATIONAL FIELDS INTERACTION:
  PARTICLE - LIKE SOLUTIONS}},
  \href{https://doi.org/10.1016/0003-4916(79)90235-5}{\emph{Annals Phys.}
  {\bfseries 118} (1979) 84--107}.

\bibitem{Yajima:2000kw}
H.~Yajima and T.~Tamaki, \emph{{Black hole solutions in Euler-Heisenberg
  theory}}, \href{https://doi.org/10.1103/PhysRevD.63.064007}{\emph{Phys. Rev.}
  {\bfseries D63} (2001) 064007},
  [\href{https://arxiv.org/abs/gr-qc/0005016}{{\ttfamily gr-qc/0005016}}].

\bibitem{Stefanov:2007zza}
I.~Z. Stefanov, S.~S. Yazadjiev and M.~D. Todorov, \emph{{Scalar-tensor black
  holes coupled to Euler-Heisenberg nonlinear electrodynamics}},
  \href{https://doi.org/10.1142/S0217732307023560}{\emph{Mod. Phys. Lett.}
  {\bfseries A22} (2007) 1217--1231},
  [\href{https://arxiv.org/abs/0708.3203}{{\ttfamily 0708.3203}}].

\bibitem{Corda:2009xd}
C.~Corda and H.~J. Mosquera~Cuesta, \emph{{Removing black-hole singularities
  with nonlinear electrodynamics}},
  \href{https://doi.org/10.1142/S0217732310033633}{\emph{Mod. Phys. Lett.}
  {\bfseries A25} (2010) 2423--2429},
  [\href{https://arxiv.org/abs/0905.3298}{{\ttfamily 0905.3298}}].

\bibitem{Ruffini:2013hia}
R.~Ruffini, Y.-B. Wu and S.-S. Xue, \emph{{Einstein-Euler-Heisenberg Theory and
  charged black holes}},
  \href{https://doi.org/10.1103/PhysRevD.88.085004}{\emph{Phys. Rev.}
  {\bfseries D88} (2013) 085004},
  [\href{https://arxiv.org/abs/1307.4951}{{\ttfamily 1307.4951}}].

\bibitem{Hendi:2014xia}
S.~H. Hendi and M.~Allahverdizadeh, \emph{{Slowly Rotating Black Holes with
  Nonlinear Electrodynamics}},
  \href{https://doi.org/10.1155/2014/390101}{\emph{Adv. High Energy Phys.}
  {\bfseries 2014} (2014) 390101}.

\bibitem{Breton:2016mqh}
N.~Breton and L.~A. Lopez, \emph{{Quasinormal modes of nonlinear
  electromagnetic black holes from unstable null geodesics}},
  \href{https://doi.org/10.1103/PhysRevD.94.104008}{\emph{Phys. Rev.}
  {\bfseries D94} (2016) 104008},
  [\href{https://arxiv.org/abs/1607.02476}{{\ttfamily 1607.02476}}].

\bibitem{Maceda:2018zim}
M.~Maceda and A.~Macías, \emph{{Non-commutative inspired black holes in
  Euler–Heisenberg non-linear electrodynamics}},
  \href{https://doi.org/10.1016/j.physletb.2018.11.048}{\emph{Phys. Lett.}
  {\bfseries B788} (2019) 446--452},
  [\href{https://arxiv.org/abs/1807.05269}{{\ttfamily 1807.05269}}].

\bibitem{Olvera:2019unw}
J.~C. Olvera and L.~A. López, \emph{{Scattering and absorption sections of
  nonlinear electromagnetic black holes}},
  \href{https://arxiv.org/abs/1910.03067}{{\ttfamily 1910.03067}}.

\bibitem{Kruglov:2017fck}
S.~I. Kruglov, \emph{{Black hole as a magnetic monopole within exponential
  nonlinear electrodynamics}},
  \href{https://doi.org/10.1016/j.aop.2016.12.036}{\emph{Annals Phys.}
  {\bfseries 378} (2017) 59--70},
  [\href{https://arxiv.org/abs/1703.02029}{{\ttfamily 1703.02029}}].

\bibitem{Kruglov:2017mpj}
S.~I. Kruglov, \emph{{Born–Infeld-type electrodynamics and magnetic black
  holes}}, \href{https://doi.org/10.1016/j.aop.2017.06.008}{\emph{Annals Phys.}
  {\bfseries 383} (2017) 550--559},
  [\href{https://arxiv.org/abs/1707.04495}{{\ttfamily 1707.04495}}].

\bibitem{Kruglov:2017xmb}
S.~I. Kruglov, \emph{{Nonlinear Electrodynamics and Magnetic Black Holes}},
  \href{https://doi.org/10.1002/andp.201700073}{\emph{Annalen Phys.} {\bfseries
  529} (2017) 1700073}, [\href{https://arxiv.org/abs/1708.07006}{{\ttfamily
  1708.07006}}].

\bibitem{Kruglov:2018rrm}
S.~I. Kruglov, \emph{{Magnetically charged black hole in framework of nonlinear
  electrodynamics model}},
  \href{https://doi.org/10.1142/S0217751X18500239}{\emph{Int. J. Mod. Phys.}
  {\bfseries A33} (2018) 1850023},
  [\href{https://arxiv.org/abs/1803.02191}{{\ttfamily 1803.02191}}].

\bibitem{Kruglov:2018lct}
S.~I. Kruglov, \emph{{On a model of magnetically charged black hole with
  nonlinear electrodynamics}},
  \href{https://doi.org/10.3390/universe4050066}{\emph{Universe} {\bfseries 4}
  (2018) 66}, [\href{https://arxiv.org/abs/1805.07595}{{\ttfamily
  1805.07595}}].

\bibitem{Novello:1999pg}
M.~Novello, V.~A. De~Lorenci, J.~M. Salim and R.~Klippert, \emph{{Geometrical
  aspects of light propagation in nonlinear electrodynamics}},
  \href{https://doi.org/10.1103/PhysRevD.61.045001}{\emph{Phys. Rev.}
  {\bfseries D61} (2000) 045001},
  [\href{https://arxiv.org/abs/gr-qc/9911085}{{\ttfamily gr-qc/9911085}}].

\bibitem{Ndongmo:2019ywh}
R.~T. Ndongmo, S.~Mahamat, T.~B. Bouetou and T.~C. Kofane, \emph{{Thermodynamic
  of a rotating and Non-linear magnetic-charged black hole in the quintessence
  field}},  \href{https://arxiv.org/abs/1911.12521}{{\ttfamily 1911.12521}}.

\bibitem{DeLorenci:2000yh}
V.~A. De~Lorenci, R.~Klippert, M.~Novello and J.~M. Salim, \emph{{Light
  propagation in nonlinear electrodynamics}},
  \href{https://doi.org/10.1016/S0370-2693(00)00522-0}{\emph{Phys. Lett.}
  {\bfseries B482} (2000) 134--140},
  [\href{https://arxiv.org/abs/gr-qc/0005049}{{\ttfamily gr-qc/0005049}}].

\bibitem{Bardeen:1973tla}
J.~M. Bardeen, \emph{{Timelike and null geodesics in the Kerr metric}},  in
  \emph{{Proceedings, Ecole d'Eté de Physique Théorique: Les Astres Occlus:
  Les Houches, France, August, 1972}}, pp.~215--240, 1973.

\bibitem{Blandford:1977ds}
R.~D. Blandford and R.~L. Znajek, \emph{{Electromagnetic extractions of energy
  from Kerr black holes}},
  \href{https://doi.org/10.1093/mnras/179.3.433}{\emph{Mon. Not. Roy. Astron.
  Soc.} {\bfseries 179} (1977) 433--456}.

\end{thebibliography}\endgroup

\end{document}